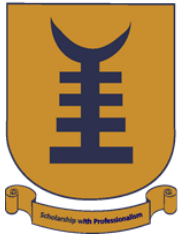
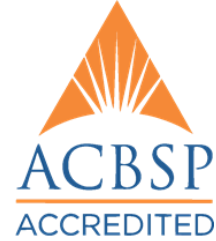

# UNIVERSITY OF PROFESSIONAL STUDIES, ACCRA (UPSA)

## PREDICTING INSURANCE PENETRATION RATE IN GHANA USING THE AUTOREGRESSIVE INTEGRATED MOVING AVERAGE (ARIMA) MODEL.

**BY**

**GROUP 3**

**THOMAS ASANTE GYIMA-ADU**

**GODWIN GIDISU**

**A DISSERTATION SUBMITTED TO THE FACULTY OF ACCOUNTING AND FINANCE OF THE UNIVERSITY OF PROFESSIONAL STUDIES, ACCRA (UPSA) IN PARTIAL FULFILMENT OF THE REQUIREMENTS FOR A BSc. (HONS.) ACTUARIAL SCIENCE DEGREE**

**MAY 2023**

# CANDIDATES' DECLARATION

We the undersigned do hereby declare that this dissertation is the result of our original research and that this work has not been previously submitted and approved by this or any other university. To the best of our knowledge and conviction, the report contains no material previously published or written by another person except where reference is made in report itself.

| NAME OF STUDENT | INDEX NUMBER | SIGNATURE | DATE |
|---|---|---|---|
| **GYIMA-ADU THOMAS** | **10114770** | ……………. | …………… |
| **GIDISU GODWIN** | **10111654** | …………… | ….............. |

# SUPERVISOR'S DECLARATION

I declare that the preparation of this dissertation was in accordance with the guidelines on supervision of dissertation laid down by the University of Professional Studies, Accra (UPSA).

SUPERVISOR'S NAME

DR. KOJO ESSEL MENSAH

SIGNATURE

………………………..

# ABSTRACT

Ghana records a low penetration rate of 1.05 % compared to some of its African counterparts. For example, South Africa, which has an insurance penetration rate of 17% followed by Namibia, which records 6.3%. This means, there is more room for improvement. More upsetting, with Ghana hovering around 1% as at 2018, this rate works out to the small amount of Gross Domestic Product. Ghana's relatively low premium are partially responsible for the country's muted penetration; for example, life insurance has a coverage rate of 32%, according to the NIC. This research seeks to model and forecast insurance penetration rate in Ghana using the autoregressive integrated moving average technique. The result indicates that ARIMA (3,1,0) is appropriate model for the insurance penetration rate in Ghana. Also, results from the forecast could serve as an advisory or the need to re-strategize as a country at the right time. Therefore, determining the future pattern of insurance penetration will lead to the remedies that will increase the number of insured in the future.

# DEDICATION

This dissertation is dedicated to our beloved parents and benefactors for financing our education and also to all UPSA teaching and non-teaching staff for their explicit support and most importantly to Almighty God for walking this journey with us.

# ACKNOWLEDGEMENTS

We would like to acknowledge the Almighty God for being the source of our inspiration, good health and being with us through our entire journey through the University of Professional Studies, Accra.

We acknowledge our families for their prayers and for always being our cheerleaders.

To our unrelenting supervisor, Dr Kojo Essel Mensah under whose guidance and supervision this research has become a success, we are extremely grateful.

We further extend our gratitude to Dr Albert Ayi Ashiagbor whose contribution especially in the area of Data and Machine learning, has led to our successful completion of the study.

# TABLE OF CONTENTS

# LIST OF ABBREVIATIONS AND ACRONYMS

| | | |
|---|---|---|
| ACF | - | Autocorrelation Function |
| ADF Test | - | Augmented Dickey Fuller Test |
| AIC | - | Akaike Information Criterion |
| AICc | - | Corrected Akaike Information Criterion |
| AR | - | Autoregressive |
| ARIMA | - | Autoregressive Integrated Moving Average |
| BIC/SIC | - | Bayesian Information Criterion/ Schwarz Information Criterion |
| GDP | - | Gross Domestic Product |
| GSS | - | Ghana Statistical Service |
| IAIS | - | International Association of Insurance Supervisors |
| IP | - | Insurance Penetration |
| IPR | - | Insurance Penetration Rate |
| IPRforecast | - | Insurance Penetration Rate Forecast (Variable) |
| IPRmodel | - | Insurance Penetration Rate Model (Variable) |
| IPRtseries | - | Insurance Penetration Rate Time Series (Variable) |
| KPSS Test | - | Kwiatkowski-Phillips-Schmidt-Shin Test |
| MA | - | Moving Average |
| MAE | - | Mean Absolute Error |
| MAPE/MAPD | - | Mean Absolute Percentage Error/ Mean Absolute Percentage Deviation |
| MASE | - | Mean Absolute Scaled Error |
| MSE/MSD | - | Mean Squared Error/ Mean Squared Deviation |
| NIC | - | National Insurance Commission |
| PACF | - | Partial Autocorrelation Function |
| PP Test | - | Phillips-Perron Test |
| RMSE/RMSD | - | Root-Mean-Square Error/ Root-Mean-Square Deviation |
| s.e | - | Standard Error |
| SSE | - | Sum of Squared Estimates of Errors/Residuals |

# DEFINITION OF TERMS

| | |
|---|---|
| **Bancassurance** | Bancassurance is an arrangement between a bank and an insurance company, through which the insurer can sell its products to the bank's customers. |
| **Insurance penetration rate** | Total insurance premiums as a percentage of gross domestic product (GDP) |
| **Insurance density** | Premiums per capita (in USD) |
| **Microinsurance** | Low-income segment of the insurance market that offer simple products, procedures, policies and low premium. |
| **Microinsurance penetration rate** | Microinsurance premiums as a percentage of GDP |
| **Microinsurance density** | Microinsurance premiums per capita (in USD) |
| **R** | A programming language for statistical computing |
| **RStudio** | An integrated development environment (IDE) for R. |
| **Time Series** | A time series is a sequence of ordered data. The "ordering" refers generally to time, but other orderings could be envisioned (e.g., over space, etc) |

# LIST OF EQUATIONS

# LIST OF FIGURES

# LIST OF TABLES

# CHAPTER ONE

# 1 INTRODUCTION

## 1.1 Background

Insurance penetration rate is an important measure of the insurance industry's performance in each country. It refers to the percentage of the population that has insurance coverage against potential risks, such as health, property damage, and loss of income. A measure of the development of an insurance sector is insurance penetration, defined as gross premium income (GPI) as a percentage of gross domestic product (GDP). According to Mahul et al (2009), insurance penetration rate is expressed as the ratio between insurance premium volume and GDP; non- life insurance penetration is expressed as the ratio between non- life insurance premium volume and GDP. Saunders & Cornett (2008) point out that Insurance serves a number of valuable economic functions that are largely distinct from other types of financial intermediaries. To highlight the unique characteristics of insurance, it is worth highlighting services that other financial service providers do not offer excluding for instance the contractual savings features of whole or universal life products. The indemnification and risk pooling properties of insurance facilitate commercial transactions and the provision of credit by mitigating losses as well as the measurement and management of non-diversifiable risk more generally. Most fundamentally, the availability of insurance enables risk adverse individuals and entrepreneurs to undertake higher risk, higher return activities than they would do in the absence of insurance, promoting higher productivity and growth.

In Ghana, the insurance industry has been growing steadily in recent years, but the insurance penetration rate remains low compared to other countries in the region. To address this issue, researchers have turned to predictive modelling techniques to forecast future trends in insurance penetration rates. One such technique is the Autoregressive Integrated Moving Average (ARIMA) model, which is a time series analysis method that uses historical data to make predictions about future values.

The ARIMA model has been widely used in various fields, including finance, economics, and healthcare, to predict future trends and patterns. It is particularly useful in forecasting time-dependent data, such as monthly or yearly insurance penetration rates in Ghana. By analysing past trends, seasonality, and other factors that influence insurance coverage, the ARIMA model can provide insights into future trends that can inform policy and business decisions.

Overall, the use of the ARIMA model to predict insurance penetration rates in Ghana can help stakeholders in the insurance industry and policymakers to make informed decisions about how to improve coverage and increase public awareness about the importance of insurance.

Insurance Act 2021 (Act 1061) governs the Insurance Industry. This Act complies significantly with the International Association of Insurance Supervisors (IAIS) Core Principles and gives better regulatory powers to the National Insurance Commission (NIC). The Act among other things prohibits composite insurance companies. All composite insurance companies therefore had to separate their life and non-life operations into different companies by December 2007.

NIC is given a broad consumer protection and prudential regulation mandate by Section 2(3) of the Act, which states that in performing its functions under the Act, the Commission shall have regard to the protection of the public against financial loss arising out of the dishonesty, incompetence, malpractice or insolvency of insurers or insurance intermediaries. The Act, among other things, prohibits composite insurance companies. All composite insurance companies, therefore, had to separate their life and non-life operations into different companies.

A study by the National Insurance Commission shows that about 30% of the Ghanaian population is covered by insurance. The total profit that was declared by the insurance industry by the end of 2018 was GHS 202 million and that of total corporate tax was GHS 36 million. Total premium for 2018 in the insurance industry was given as GHS 2,937,534,716. The market is young with 29 non-life insurance companies, 24 life insurance companies, 3 reinsurance companies, 82 Brokerage companies, 4 Reinsurance Brokerage firms and 3 insurance loss adjusters competing in the industry as of December 2018. Insurance penetration as a percentage to GDP in Ghana was below 1.85% at the end of first quarter of 2016 and 1.2% in the last quarter of 2017, there is a general consensus that, the sector is underperforming.

The growth of insurance may have a direct bearing on economic development, creating value, and sustainability for all stakeholders in the insurance business (Ghosh, 2013; Akinlu & Apansile, 2014).

It is important to state that, there are differences in the calculation of insurance premium among different countries. In the case of Ghana, the calculation excludes pension and health insurance as this is not the case for some countries. The National Insurance Commission does not regulate pension and health insurance. Therefore, for the purposes of our research, we shall forecast insurance premiums based on data provided by the National Insurance Commission, which excludes pensions and health insurance.

## 1.2 Problem Statement

Whereas access to financial services is a potentially important means of alleviating poverty, especially when combined with other supports for poor households, yet access to insurance products has yet to take off in most segments of the Ghanaian population, reaching only a small segment of the potential market as indicated by low penetration levels. Although long neglected by mainstream financial firms, it would be a mistake to think that Insurance requires some special alchemy for its functioning.

Arena, M. (2006) argues that the deepening of insurance and banking systems appears to play a complementary role in the growth process. Insurance companies and banks each make a positive contribution to growth, but each contributes more when both are present. The development of the insurance market also contributes to the soundness of the securities market. The deepening of the insurance market has made a positive contribution to economic growth.

Life insurance is causal to growth only in high-income countries, while non-life insurance makes a positive contribution in both emerging and high-income countries. Some research suggests that the positive contribution of life insurance to growth comes mainly from financial intermediation and long-term investments. The study points to a positive relationship between insurance penetration and economic growth, but in developing countries, it is trivial and as a result does not lead to economic growth. The aim of the study is to predict insurance penetration rates using the Autoregressive Integrated Moving Average model.

Most of the theoretical queries and insurance penetration modes have centred on individual developed countries such as the U.S. or have been advanced across continent-wide queries. Less exposure is paid to developing countries like Ghana in the attempt to forecast insurance premiums hence insurance penetration rate in Ghana. There is currently no linear forecasting models and macroeconomic theory models for insurance penetration rate forecasting in Ghana. Previous research has generalized the financial sector and put more emphasis on into the banking subsector.

However, these studies do not exhaustively address the contribution of insurance industry penetration to economic growth, as the insurance industry is an important subsector of the financial sector. The research conducted evaluates the complementary role played by the insurance and banking finance sub-sectors in impact on economic growth. The banking subsector has gained more prominence at the expense of other equally important financial subsectors.

Understanding the factors that influence insurance penetration rates and predicting their future trends can help insurance companies and policymakers develop effective strategies to increase insurance coverage and promote financial security.

## 1.3 Purpose of the Study

The purpose of the study is in twofold. First, to ascertain which econometric model for time series analysis fits the Ghanaian insurance penetration rate most accurately and further do forecast to validate the fit. Secondly, this study aims to undertake a comprehensive analysis of GDP forecasting and modelling for a developing country which could be used further as a template for individuals who are interested in undertaking ambitious research of modelling a country's insurance penetration rate.

Our contribution to the literature is the claim that is the first study analyzing the quarterly data for Ghana's insurance penetration rate series to estimate and forecast the data process.

This paper aims to forecast insurance penetration in Ghana, acknowledging variables that lead to insurance penetration such as insurance premiums and Gross Domestic Product (GDP).

Lessons can be taken by distinguishing variables that lead to the consumption of insurance and how to enhance these variables in order to grow an insurance market that can promote economic growth.

This study therefore will offer an insight into the challenges and strategies to increase insurance penetration concerning all the various types of insurance companies.

## 1.4 Objectives of the Study

### 1.4.1 General objective

The main objective of this research is to predict insurance penetration rates in Ghana using ARIMA and discover solutions to increase insurance penetration in Ghana.

### 1.4.2 Specific Objectives

i) To analyse the historical trend of insurance penetration rates in Ghana

ii) To develop an ARIMA model to forecast insurance penetration rates in Ghana.

iii) Evaluate the accuracy of the ARIMA model selected and compare it with other ARIMA models.

iv) To suggest strategies to increase insurance penetration in Ghana.

## 1.5 Research Questions

i) What does the historical trend of insurance penetration rate in Ghana portray?

ii) What ARIMA model is best to forecast insurance penetration rate in Ghana?

iii) How accurate is the ARIMA model selected compared to other ARIMA models?

iv) How can insurance penetration in Ghana be increased?

## 1.6 Significance of the Study

This study will be of interest to regulators i.e., the National Insurance Commission and financial system institutions, economic analysts and other subjects in Ghana in order to carry out a timely analysis and forecast of the trend in insurance market growth.

Insurance penetration infers the contribution of the insurance sector to the GDP and reveals the development of the sector. Therefore, should inform the economy-wide policies developed by the government and the regulators of insurance markets in these economies. This is necessary because all these policies will promote or mar the growth of the insurance sector. Furthermore, our research will add to existing literature on Insurance Penetration as well and Time Series Forecasting. Our findings could be of relevance in finance related modules in higher education institutions and should not be restricted to only commerce courses. This will lead to the exposure of students to the benefits of financial services in particular insurance. Moreover, the research and development segment of insurance seeks to investigate the viability of growth opportunities in the insurance sector. This will ensure that growth opportunities are maximised accurately and efficiently.

This study will propose using the same or comparable models to predict the Consumer Price Index and Gross Domestic Product (GDP) in different sectors of the financial sector. Because ARIMA are versatile and may be used with different variables without changing the structure of the model, this will allow for additional use of the model in the same environment. Time series is a general issue that has significant practical value in many fields, including hydrology, finance, and economics.

## 1.7 Scope of the Study

The study is mainly focused ARIMA for Insurance penetration rate forecasting inclusive of predictors as discussed in detail in this study. The model focused on incorporating Holt-Winters exponential smoothing in insurance penetration rate in Ghana. The data used to test the model were collected from secondary data sources.

# CHAPTER TWO

## 2 LITERATURE REVIEW

### 2.1 Introduction

Forecasting the future state of a certain topic of interest is vital when making decisions. In order to forecast potential future outcomes for the study, one must first identify inputs that are essential to the prediction process. These predictions mainly rely on historical data and the assumption that past variable behaviour will replicate itself in the future.

A good forecast model is thus defined by its reliability, ease of use, having an output that is meaningful, compatibility with other systems, timeliness of the forecast and reliable accuracy. (Arienda, Asana & Costantino 2015).

As the time horizon gets further into the future, forecasts become distorted, and no single forecasting method can be used exclusively. The choice of a forecasting model is guided by a set of criteria.

Most forecasting methods require the use of a computer system with specific types of software to run the predictor of choice. In our dissertation, we used RStudio for forecasting.

This chapter presents a review of relevant theoretical and empirical literature on time series forecasting and its resulting application in forecasting insurance penetration rates in Ghana. The theoretical review presents the theoretical background to insurance penetration in Ghana. The second part reviews empirical studies on the penetration of insurance in Ghana. Insights from the previous sections are organized into a conceptual framework in the last section.

## 2.2 Theoretical Review

This theoretical literature review aims to examine the theoretical underpinnings of using the ARIMA model in predicting insurance penetration rate.

The ARIMA model is based on the Box-Jenkins approach, which involves three steps: identification, estimation, and diagnostic checking. The identification step involves identifying the order of the ARIMA model based on the autocorrelation function (ACF) and partial autocorrelation function (PACF) plots. The estimation step involves estimating the parameters of the ARIMA model using maximum likelihood estimation. The diagnostic checking step involves checking the adequacy of the model by examining the residuals for autocorrelation, heteroscedasticity, and normality.

The ARIMA model is particularly useful in time series analysis because it considers the autocorrelation and trends in the data. Autocorrelation occurs when a variable is correlated with its own past values. Trend is the systematic pattern in the data over time, which can be linear or nonlinear.

The ARIMA model can be used to predict future values of insurance penetration rate by analysing the historical data on insurance penetration rate. The model can help identify the trend in insurance penetration rate, such as whether it is increasing, decreasing, or stable. The model can also identify the autocorrelation in the data, which can help predict future values of insurance penetration rate based on past values.

The ARIMA model is a powerful tool for predicting insurance penetration rate. The model considers the autocorrelation and trends in the data, which are important factors in predicting future values of insurance penetration rate. The ARIMA model can help insurance companies, policy makers, and researchers develop effective strategies to improve insurance penetration rate. However, it is important to note that the ARIMA model is not a perfect tool and should be used in conjunction with other methods to make accurate predictions.

**2.3     Empirical Review**

Empirical literature reviews are knowledge derived from investigation, observation experimentation, or experience. This literature review aims to examine the empirical literature on predicting insurance penetration rate using the ARIMA model.

Several studies have been conducted on predicting insurance penetration rate using the ARIMA model. This review focuses on studies that have used the ARIMA model to predict insurance penetration rate in different countries.

Raymond Y.C. Tse, (1997) suggested that the following two questions must be answered to identify the data series in a time series analysis: (1) whether the data are random; and (2) have any trends? This is followed by another three steps of model identification, parameter estimation and testing for model validity. If a series is random, the correlation between successive values in a time series is close to zero. If the observations of time series are statistically dependent on each another, then the ARIMA is appropriate for the time series analysis.

Meyler et al (1998) drew a framework for ARIMA time series models for forecasting Irish inflation. In their research, they emphasized heavily on optimizing forecast performance while focusing more on minimizing out-of-sample forecast errors rather than maximizing in-sample 'goodness of fit'.

Stergiou (1989) in his research used ARIMA model technique on a 17 years' time series data (from 1964 to 1980 and 204 observations) of monthly catches of pilchard (Sardina pilchardus) from Greek waters for forecasting up to 12 months ahead and forecasts were compared with actual data for 1981 which was not used in the estimation of the parameters. The research found mean error as 14% suggesting that ARIMA procedure was capable of forecasting the complex dynamics of the Greek pilchard fishery, which, otherwise, was difficult to predict because of the year-to-year changes in oceanographic and biological conditions.

A study by Mishra and Dash (2012) used the ARIMA model to predict insurance penetration rate in India for the period 2012 to 2016. The results showed that insurance penetration rate was predicted to increase from 5.15% in 2012 to 5.91% in 2016. The study also found that income level, level of education, and awareness of insurance were significant predictors of insurance penetration rate in India.

Nwolley-Kwasi (2015) employs the ARIMA technique to forecast the growth pattern of claims payment. The researchers used the Suitable Box-Jenkins model to examine time-series data from January 2010 to November 2014. The results demonstrate that the ARIMA(1,1,0) model performed the best in terms of predicting claim payment.

Namawejje and Geofrey (2020) used yearly data from 2000 to 2018 to forecast life insurance premiums and insurance penetration rates. The best model was ARIMA (0, 1, 0). The findings show a slightly increasing trend of 0.9% to 1.19 % favouring individual life premiums during the period of forecasting. Deposit administration, on the other hand, as well as group life premiums, remained constant.

Yang et al. (2016) developed an ARIMA model to forecast China's GDP. Using the ARIMA (2, 4, 2) model derived from EViews 6.0, their results showed that China's GDP reached 72,407.76 billion yuan in 2016 and 77,331.48 billion yuan in 2017. They also forecasted GDP for the next two years, which showed an increasing trend, owing to China's shifting economic growth strategy at the time of transition from an investment-driven to an innovation-driven approach.

The empirical literature suggests that the ARIMA model can be used to predict insurance penetration rate in different countries with reasonable accuracy. The results of the studies reviewed here suggest that income level, level of education, awareness of insurance, economic growth, and regulatory environment are significant predictors of insurance penetration rate. These findings provide useful insights for insurance companies to develop effective marketing strategies and improve their operations to increase insurance penetration rate in different countries.

### 2.3.1   Attempts at Insurance Penetration Rate Forecasting in Ghana and other African countries.

Ghana plays a significant role in the West African region and records the highest total insurance premium in the region as of 2018. In Ghana, there have been no attempt to forecast insurance penetration in Ghana.

Most studies focused on analysing the long- and short-term dynamics of the determinants of insurance, of which insurance penetration rate, non-life insurance penetration rate and life insurance penetration rates where variables used in the study.

Alhassan and Fiador (2014) used autoregressive distributed lag bounds to find a long-run positive relationship between insurance penetration and economic growth which implies that funds mobilized from insurance businesses have a long-run impact on economic growth. The model is represented by,

*Equation 2.1 Autoregressive Distributed Lag (ARDL) framework*

$$\Delta growth_t = \alpha_0 + \sum_{i=1}^{p} \theta_i \Delta growth_{t-1} + \sum_{i=0}^{p} \mu_i \Delta \ln ip_{t-1} + \sum_{i=0}^{p} \pi_i \Delta \ln gcap_{t-1} + \sum_{i=0}^{p} \delta_i \Delta inf_{t-1} + \sum_{i=0}^{p} \varphi_i \Delta \ln trd_{t-1} + \lambda_1 \ln ip_{t-1} + \lambda_2 gcap_{t-1} + \lambda_3 inf_{t-1} + \lambda_4 trd_{t-1} + \varepsilon_t$$

Olarewaju and Msomi (2021) in their study analysed the long- and short-term dynamics of the determinants of insurance penetration for the period 1999 Q1 to 2019 Q4 in 15 West African countries. Panel auto regressive distributed lag (P-ARDL) model was used on the quarterly data gathered A co-integrating and short-run momentous connection was discovered between insurance penetration along with the independent variables, which were education, productivity, dependency, inflation and income.

The model in P-ARDL format is given by,

*Equation 2.2 Model in P-ARDL format*

$$\Delta INP_{it} = q_0 + \sum_{g=1}^{n} \beta_{1g} \Delta INP_{it-g} + \sum_{g=1}^{n} \beta_{2g} \Delta INF_{it-g} + \sum_{g=1}^{n} \beta_{3g} \Delta EDU_{it-g} + \sum_{g=1}^{n} \beta_{4g} \Delta \log PRO_{it-g} + \sum_{g=1}^{n} \beta_{5g} \Delta DEP_{it-g} + \sum_{g=1}^{n} \beta_{6g} \Delta \log INC_{it-g} + \forall_1 INP_{it-1} + \forall_2 INF_{it-1} + \forall_3 EDU_{it-1} + \forall_4 \log PRO_{it-1} + \forall_5 DEP_{it-1} + \forall_6 \log INC_{it-1} + u_{it}$$

INP is the variable for insurance penetration, INF for inflation rate, EDU for education level, PRO for the productivity of labour, DEP for dependency ratio and INC for income level where Δ signifies the first difference operator, *g* signifies the selected number of lags and *n* signifies the optimal lag length; $q_0$ is the constant term and $u_{it}$ is the composite error term.

$\beta_{1g}$–$\beta_{6g}$ are the short-run coefficients of the regressors, respectively, while $\forall_1$–$\forall_6$ are the long-run coefficients of the regressors. *i* represents the number of countries (15) examined, while t represents the number of years examined.

### 2.3.2 Attempts at Insurance Penetration Rate Forecasting in Nigeria

Hafiz, Salleh, Garba, and Rashid (2021) used annual data on insurance penetration (IP) for the period 1981 to 2018 to project insurance penetration rate for 12 years (2019 to 2030) with the ARIMA model. Their results showed that insurance penetration rate keeps decreasing but at a slower rate. The model eventually selected was the ARIMA (2, 0, 1) which is specified as below:

*Equation 2.3 ARIMA (2, 0, 1) Model*

$$IP_1 = c + \beta_2 IP_{t-2} + \alpha_2 \mu_{t-2} + \alpha_1 \mu_{t-1} + \mu_t$$

## 2.4 Insurance Penetration Rate Forecast Techniques

Forecast techniques can be identified as casual, time series and smoothing techniques. These forecasting techniques are data intensive and classified as an illustration as shown in the figure below.

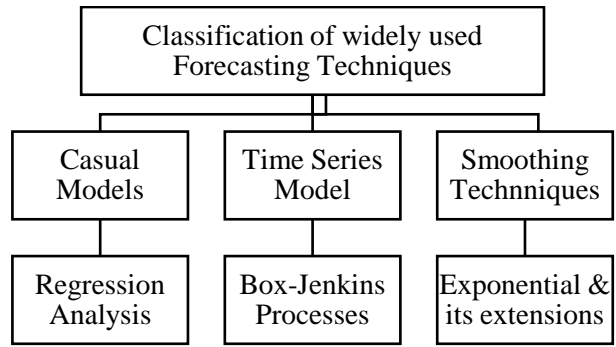

*Figure 2.1 Classification of forecasting techniques (Arsham, 2006)*

Forecasting techniques can also be categorized into judgemental, consumer/market research, cause-effect and artificial intelligence as shown below,

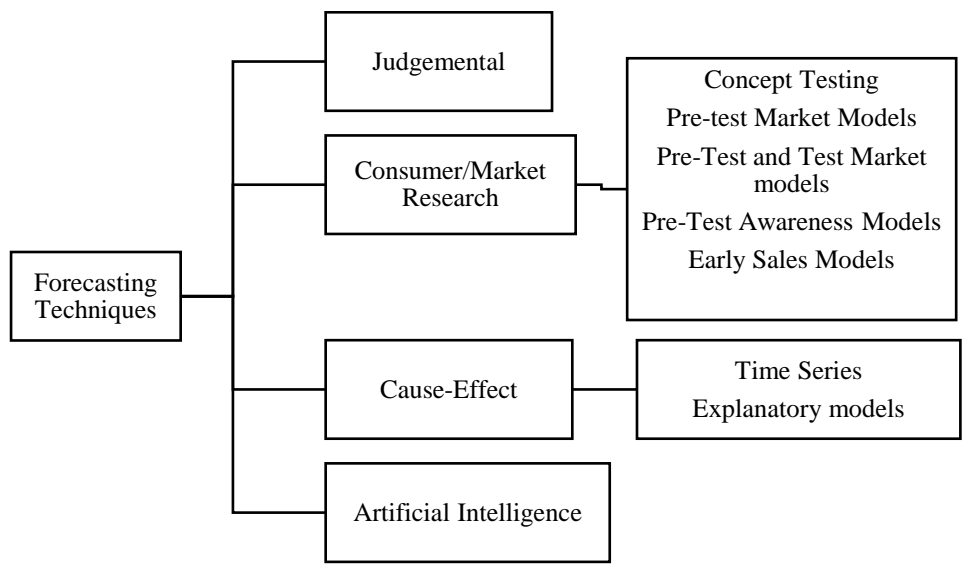

*Figure 2.2 Classification of Forecasting Techniques (Webby & O'Connor, 2014)*

# CHAPTER THREE

## 3 RESEARCH METHODOLOGY

### 3.1 Introduction

An autoregressive integrated moving average (ARIMA) model was used to predict a lot of data. These models were used because each observation explains only one variable. Scholars from different disciplines, such as Jose and Sojan (2013), Yeboah, Ohene and Wereko (2012), Wang and Yang (2017), Anand Madhu (2014), Mandal (2005), Raymond (1997), have used these models' prediction. A key assumption of these models is that in time series analysis there is an aspect that past patterns persist in the future (Ramasubramanian, 2007; Heng, Zhang & Yang, 2017). These models capture patterns and use them to predict future expected values.

### 3.2 Data Collection

The data sample employed quarterly data for life insurance premiums and insurance penetration rates for the period from 2013 to 2022. Secondary data used is obtained from the National Insurance Commission (NIC) and Ghana Statistical Service (GSS).

We used Non-Oil GDP obtained from the Ghana Statistical Service and life and non-life gross insurance premiums obtained from National Insurance Commission. The Ghana Oil and Gas Insurance Pool (GOGIP) premium was not included in the non-life insurance gross premiums. The purpose of excluding COGIP is to achieve homogeneity in measuring both GDP and total Gross Premium. That is to exclude the effects from the Oil and Gas Sector.

Total quarterly Gross Insurance Premium was divided by total gross GDP to give quarterly insurance penetration rate in percentages.

**3.3    Model Description**

ARIMA models are divided into three components based on data type. The first component is an autoregressive (AR) time series model, which consists of past observations of the dependent variable (that is, the variable of interest) in predicting future observations.

The second component is the moving average (MA) model, which incorporates previous observations (that is, previous prediction errors) in a white noise process to predict future observations of the dependent variable.

Combining the two models MA and AR results in the stationary model ARMA. If the data used are nonstationary, a third component is used to transform the data to achieve stationarity by differencing (integrating (I)) the original series presented according to Rohrbach and Kiriwaggulu (2001) and Nau (2018)

**3.3.1    Box Jenkins Methodology**

Box Jenkins Methodology In time series analysis, the Box-Jenkins method, named after the statisticians George Box and Gwilym Jenkins, applies autoregressive moving average (ARMA) or autoregressive integrated moving average (ARIMA) models to find the best fit of a time series model to past values of a time series. The following are steps for Box-Jenkins methodology:

- ➢ Plot the series and identify the trend (is the series trending in the mean/Variance?)
- ➢ Test for stationarity (Augmented Dickey Fuller Test, KPSS Test, and PP-Test)
- ➢ Transform the series into a stationary series (power transformation /seasonal differencing/ first differencing etc.)

- Plot the Autocorrelations and Partial Autocorrelation functions (Identify possible models)
- Estimate the possible models (check for coefficient significance and white noise residuals)
- Select the best models based on the information criteria (the model with lowest AIC and BIC )

### 3.3.2 Tests for Stationarity

#### 3.3.2.1 Augmented Dickey-Fuller Test

Said and Dickey (1984) augment the basic autoregressive unit root test to accommodate general ARMA (p,q) models with unknown orders and their test is referred to as the augmented Dickey-Fuller (ADF) test.

To test for a unit root we assume that

$$\phi_p(B) = (1 - B)\varphi_{p-1}(B)$$

*Equation 3.1*

Where $\varphi_{p-1}(B) = 1 - \varphi_1 B - \cdots - \varphi_{p-1} B^{p-1}$ has roots lying outside the unit circle

$$Y_t = \phi Y_{t-1} + \sum_{j=1}^{p-1} \varphi_j \Delta Y_{t-j} + \theta_0 + \varepsilon_t$$

*Equation 3.2 ADF Test Equation*

Hypothesis

$$H_0: \phi = 1$$

$$H_1: |\phi| < 1$$

Hence reject $H_0$ if $t_{\varphi=1}$ < Critical Value (CV). The series should be differenced.

### 3.3.2.2 Phillips-Perron Test

- Phillips and Perron (1988) have developed a more comprehensive theory of unit root non-stationarity. The tests are similar to ADF tests. The Phillips-Perron (PP) unit root tests differ from the ADF tests mainly in how they deal with serial correlation and heteroscedasticity in the errors. In particular, where the ADF tests use a parametric autoregression to approximate the ARMA structure of the errors in the test regression, the PP tests ignore any serial correlation in the test regression.

- The tests usually give the same conclusions as the ADF tests, and the calculation of the test statistics is complex.

- Consider a model

$$Y_t = \theta_0 + \phi Y_{t-1} + \varepsilon_t$$

*Equation 3.3*

DF: $a_t$ ~ iid

PP: $a_t$ ~ serially correlated

$$PP\ test\ equation: \Delta Y_t = \theta_0 + \varphi Y_{t-1} + \varepsilon_t$$

*Equation 3.4 Phillips-Perron Test Equation*

Add a correction factor to the DF test statistic. (ADF is to add lagged $\Delta Y_t$ to 'whiten' the serially correlated residuals)

The hypothesis to be tested

$$H_o: \varphi = 0$$

$$H_1: \varphi < 0$$

The PP tests correct for any serial correlation and heteroscedasticity in the errors $\varepsilon_t$ of the test regression by directly modifying the test statistics $t_{\delta=0}$ and    .

Under the null hypothesis that $\delta = 0$, the PP $Z_t$ and $Z_\delta$ (computed by R) statistics have the same asymptotic distributions as the ADF t-statistic and normalized bias statistics

### 3.3.2.3 Kwiatkowski-Phillips-Schmidt-Shin (KPSS) Test

- To be able to test whether we have a deterministic trend vs stochastic trend, we are using KPSS (Kwiatkowski, Phillips, Schmidt and Shin) Test (1992).

- STEP 1: Regress $Y_t$ on a constant and trend and construct the OLS residuals $e=(e_1,e_2,…,e_n)'$.

- *STEP 2: Obtain the partial sum of the residuals.*

$$S_t = \sum_{i=1}^{t} \varepsilon_t$$

*Equation 3.5 KPSS Test- Sum of Residuals*

- *STEP 3: Obtain the test statistic*

$$KPSS = n^{-2} \sum_{t=1}^{n} \frac{S_t^2}{\hat{\sigma}^2}$$

*Equation 3.6 KPSS Test Statistic*

- *where $\hat{\sigma}^2$ is the estimate of the long-run variance of the residuals.*

- STEP 4: Reject $H_0$ when KPSS is large, because that is the evidence that the series wander from its mean.

### 3.3.3 Autoregressive (AR) Model

An autoregressive model of order p, AR (p), can be expressed as:

$$Y_t = \phi_0 + \phi_1 Y_{t-1} + \phi_2 Y_{t-2} + \cdots + \phi_p Y_{t-p} + \varepsilon_t$$

*Equation 3.7 Autoregressive (AR) Model of Order p*

where $\varepsilon_t$ is the error term in the equation; where $\varepsilon_t$ a white noise process, a sequence of independently and identically distributed (i.i.d) random variables with $E(\varepsilon_t) = 0$ and $var(\varepsilon_t) = \sigma^2$; i.e. $\varepsilon_t \sim iidN(0, \sigma^2)$. In this model, all previous values can have additive effects on this level $Y_t$ and so on; so, it's a long-term memory model.

### 3.3.4 Moving Average (MA)

Moving-average (MA) model A time series *{Y}* is said to be a moving-average process of order q, MA *(q)*, if:

$$Y_t = \theta_1 \varepsilon_{t-1} + \theta_2 \varepsilon_{t-2} + \cdots + \theta_q \varepsilon_{t-q} + \varepsilon_t$$

*Equation 3.8 Moving Average (MA) of Order q*

This model is expressed in terms of past errors as explanatory variables. Therefore, only $q$ errors will have effect on $Y$, however higher order errors don't have effect on $Y_t$; this means that it is a short memory model.

### 3.3.5 Autoregressive Moving Average

Autoregressive moving-average (ARMA) model A time series *{Y}* is said to follow an autoregressive moving-average process of order *p* and *q*, ARMA *(p, q)*, process if:

$$Y_t = \phi_0 + \phi_1 Y_{t-1} + \phi_2 Y_{t-2} + \cdots + \phi_p Y_{t-p} + \theta_1 \varepsilon_{t-1} + \theta_2 \varepsilon_{t-2} + \cdots + \theta_q \varepsilon_{t-q}$$

*Equation 3.9 Autoregressive Moving Average ARMA (p,q) process*

ARMA models can be further extended to nonstationary series by allowing differencing of data series resulting in ARIMA model non-seasonal model is known as ARIMA *(p, d, q)*.

### 3.3.6 Autoregressive Integrated Moving Average (ARIMA)

Here we use three parameters. *p* is the autoregressive order; *d* is the degree of differencing and *q* is the moving average order. The general form of an ARIMA model *(p,d,q)* is

$$\Delta^d Y_t = \sum_{i=1}^{p} \phi_i \Delta^d y_{t-i} + \sum_{j=1}^{q} \theta_j \varepsilon_{t-j} + \varepsilon_t$$

$$\varepsilon_t \sim N(0, \sigma^2)$$

*Equation 3.10 Autoregressive Integrated Moving Average*

$$Y'_t = \phi_0 + \phi_1 Y'_{t-1} + \phi_2 Y'_{t-2} + \cdots + \phi_p Y'_{t-p} + \theta_1 \varepsilon_{t-1} + \theta_2 \varepsilon_{t-2} + \cdots + \theta_q \varepsilon_{t-q}$$

*Equation 3.11 First Order Autoregressive Integrated Moving Average (ARIMA)*

Where $Y'_t = Y_t - Y_{t-1}$

*Equation 3.12 ARIMA differencing of the first order*

$$Y''_t = (Y_t - Y_{t-1}) - (Y_{t-1} - Y_{t-2})$$

*Equation 3.13 ARIMA differencing of second order*

$$Y''_t = Y_t - 2Y_{t-1} + Y_{t-2}$$

*Equation 3.14*

$Y_t$ is the future insurance penetration rate, $Y_t$ and $Y_{t-1}$ are the original series and lagged series, respectively.

### 3.3.6.1 Non seasonal ARIMA models

ARIMA models are models that possibly may include autoregressive (AR) terms, moving average (MA) terms and differencing (integration) operations. When differencing is required in the model it is specified as ARIMA (p, d, q), where the 'd' refers to the order of differencing, "p" is the order of autoregressive and "q" is the order of moving average. A first difference might be used to account for a linear trend in a data set as expressed in equation 3.2.

$$Y'_t = Y_t - Y_{t-1}$$

*Equation 3.15 First Order Non- Seasonal Differencing*

### 3.3.6.2 Seasonal ARIMA Models (SARIMA)

In a time series, seasonality is a regular pattern of changes that repeats over specific periods. If $s$ defines the number of time periods until the pattern repeats again, '$s$' can be define as $s = 4$ (quarters per year). It may also be days of the week, weeks of the month and so on. In a seasonal ARIMA model, seasonal AR (P) and MA (Q) terms predict $Y_t$ using data values and errors at times with lags that are multiples of $s$ (length of season). For example with quarterly data ($s = 4$), a seasonal first order AR(1) would use $Y_{t-4}$ to predict $Y_t$, and a second order seasonal AR(2) model would use $Y_{t-4}$, and $Y_{t-8}$ to predict $Y_t$. Similarly, a first order seasonal MA (1) model would use the error $\varepsilon_{t-4}$ as a predictor just as a seasonal MA (2) would use $\varepsilon_{t-4}$ and $\varepsilon_{t-8}$ for prediction. Seasonality usually causes the series to be non-stationary because of the seasonal changes in mean. This makes differencing necessary for seasonal data to achieve stationary. Seasonal differencing is defined as a difference between a value and a value with lag that is a multiple of the seasonal period "s". For instance, quarterly data ($s = 4$) will have a first seasonal difference as indicated in equation 3.7. $B$ is the backshift operator.

$$(1 - B^4)Y_t = Y_t - Y_{t-4}$$

*Equation 3.16 First Order Seasonal Differencing*

The differences from the previous year may be about the same for each quarter of the year to yield a stationary series. Seasonal differencing removes seasonal trend and can also get rid of seasonal random walk type of non-stationary. It must also be noted that when the data series has trend, non-seasonal differencing may be applied to "de-trend" the data. For this purpose, usually a first non-seasonal difference is enough to attain stationarity as below:

$$(1 - B)Y_t = Y_t - Y_{t-1}$$

*Equation 3.17*

When both seasonality and trend are present, it may be necessary to apply both a first order non-seasonal and a seasonal difference. In which case the ACF and PACF of the equation needs to be examined.

$$(1 - B^4)(1 - B)Y_t = (Y_t - Y_{t-1}) - (Y_{t-4} - Y_{t-5})$$

*Equation 3.18 First Seasonal and Non-Seasonal Differencing*

The model is written in the following notation: $ARIMA\ (p,d,q)(P,D,Q)[S]$

Where:

: Non-seasonal *AR* orders, $d$: non-seasonal differencing, $q$: non-seasonal *MA* orders, $P$: seasonal *AR* orders, $D$: seasonal differencing, $Q$: seasonal *MA* orders, $s$: seasonal period or $s = time\ span$ of repeating seasonal pattern.

## 3.4  Model Identification

First stage of ARIMA model building is to identify whether the variable, which is being forecasted, is stationary in time series or not. Stationary data are data whose statistical properties do not change over time (Studenmund, 2016). A time series is stationary if it is characterized by a constant mean and constant variance, and a time-independent autocovariance (Ramasubramanian, 2007; Studenmund, 2016). This mean, the values of variable over time varies around a constant mean and variance. If any of these properties are not met, the data are declared nonstationary. The ARIMA model cannot be built until we make this series stationary. If non-stationary data are used in the model, the results may indicate misleading relationships (Baumohl & Lyocsa, 2009). Apply the autocorrelation function (ACF) to the data to identify this problem. If the ACF plot is positive and exhibits a very slow linear decay pattern, the data are not stationary (Nau, 2018). We first have to difference the time series 'd' times to obtain a stationary series in order to have an ARIMA(*p,d,q*) model with *'d'* as the order of differencing used. Caution to be taken in differencing as over-differencing will tend to increase in the standard deviation, rather than a reduction. The best idea is to start with differencing with lowest order (of first order, $d =1$) and test the data for unit root problems. Therefore, we obtained a time series of first order differencing.

The next step is to find initial values for the seasonal and non-seasonal orders (*p* and *q*). At this step, ACF and partial ACF (PACF) are the basic analytical tools used. To compute the autocorrelation for lag k, which is referred to as a feature of stationary data, we compute the correlation between $Y_t$ and $Y_{t-k}$ at *n-k* pairs in the dataset.

$$ACF(k) = \frac{\sum(Y_t - \mu)(Y_{t-k} - \mu)}{\sum(Y_t - \mu)^2} = \frac{Cov(Y_t, Y_{t-k})}{Var(Y_t)}$$

*Equation 3.19 Autocorrelation Function (ACF)*

where $Y_t$ is the original series, $Y_{t-k}$ is the lagged version of the original series, and $\mu$ is the mean of the data (Studenmund, 2016).

PACF is defined as the linear correlation between $Y_t$ and $Y_{t-k}$, taking into account the effects of linear relationships between values at intermediate lag. The first order partial is equal to the first order autocorrelation, but the second order can be calculated as

$$PACF = \frac{Cov(Y_t, Y_{t-2}\backslash Y_{t-1})}{\sqrt{(VarY_t\backslash Y_{t-1})Var(Y_{t-2}\backslash Y_{t-1})}}$$

*Equation 3.20 Partial Autocorrelation Function (PACF)*

The ACF gives the order of AR (*p*) and PACF gives the order of MA (*q*).

The order selection of the model is very important when we use ARIMA process because the higher order in the model may result in smaller estimated errors. The Akaike Information Criterion (AIC) is a measure of the relative quality of statistical model for a given set of data, which has been used for ARIMA model selection and identification. It was used to determine if a particular model with specified parameters is a good statistical fit and the model with the lowest AIC value should be chosen. The AIC provide a researcher with an estimate of the information that would be lost if a particular model were to be used to display the process that produced the data. The Akaike showed that this criterion selects the model that minimizes:

$AIC = -2$(maximized log likelihood−number of parameters in the model).

$$AIC = -2 \log L + 2(p + q + k + 1)$$

*Equation 3.21*

For more simplified form the Akaike Information Criterion (AIC) is expressed in equation

$$AIC = \log \sigma_k^2 + \frac{n + 2k}{n}$$

*Equation 3.22  Akaike Information Criterion*

Correlated Akaike Information Criterion (AICc) is given in equation:

$$AICc = \log \sigma_k^2 + \frac{n + k}{n - k - 2}$$

*Equation 3.23 Corrected Akaike Information Criterion (AICc)*

In computing this, if a model has relatively little bias, describing reality well, it tends to provide more accurate estimates of the quantities of interest. The smaller AIC is better the model to be used.

Bayesian or Schwarz Information Criterion (BIC/SIC) is expressed in equation

$$BIC = \log \sigma_k^2 + \frac{k \log n}{n}$$

*Equation 3.24 Bayesian Information Criterion (BIC)*

Where $\sigma_k^2 = \frac{SSE(k)}{n}$ where, k is the number of parameter in the model and n is the sample size

SSE is the sum of squared estimates of errors/residuals. It is a measure of discrepancy between the data and estimation model.

## 3.5 Parameter Estimation

After identifying the appropriate orders of ARIMA(*p,d,q*), we attempt to find accurate estimates of the model's parameters using the least-squares method described by Box and Jenkins. The parameters are determined by the maximum likelihood method, which is asymptotically correct for the time series. The estimators are usually sufficient, efficient and consistent for Gaussian distributions, and asymptotically normal and efficient for some non-Gaussian families (Mandal, 2005; Nyoni & Bonga, 2019). In this study, the parameters are estimated using RStudio. Different software will give different estimates.

The log-likelihood function:

$$\ln L(\phi, \sigma_a^2) = -\frac{n}{2}\ln 2\pi - \frac{n}{2}\ln \sigma_a^2 - \frac{1}{2}\ln(1-\phi^2) - \frac{1}{2\sigma_a^2}\left[\sum_{t=2}^{n}(Y_t - \phi Y_{t-1})^2 + (1-\phi^2)Y_1^2\right]$$

*Equation 3.25 The log-likelihood function*

## 3.6 Diagnostic Checking

The estimated model must be checked to verify if it adequately represents the series. The best model was selected based on the minimum values of Root Mean Square Error (RMSE), Mean Absolute Percentage Error (MAPE), Normalized Bayesian Information Criterion (BIC), Akaike Information Criterion (AIC) and highest value of R-square. Diagnostic checks are performed on the residuals to see if they are randomly and normally distributed. Here, the Kolmogorov-Smirnov test for normality was used. An overall check of the model adequacy was made using the modified Box-Pierce $Q$ statistics. The test statistics is given by:

$$Q_m = n(n+2) \sum_{k=1}^{n} (n-k)^{-1} r_k^2 \approx \chi_{m-r}^2$$

*Equation 3.26 Box-Pierce Q Statistic*

Where,

$r_k^2$ = The residuals autocorrelation at lag k

$n$ = The number of residuals

$m$ = The number of time lags included in the test

when the *p*-value associated with the $Q$ is large the model is considered adequate, else the whole estimation process has to start again in order to get the most adequate model. Here all the tests were performed at the 5% level of significance.

We use three evaluation criteria to assess the performance of the model (Wayne A. Woodward, Henry L. Gray, Alan C. Elliott, 2017), respectively, are the mean squared error (MSE), the root mean square error (RMSE), MAE the mean absolute error and the mean absolute percentage error (MAPE); the formulations are detailed as follows:

$$ME = \frac{1}{n} \sum_{i=1}^{n} (Y_t - F_t)$$

*Equation 3.27 Mean Error (ME)*

$$MAPE = \frac{100\%}{n} \sum_{i=1}^{n} \left|\frac{Y_t - F_t}{Y_t}\right|$$

*Equation 3.28 Mean Absolute Percentage Error (MAPE)*

$$MAE = \frac{1}{n} \sum_{i=1}^{n} |Y_t - F_t|$$

*Equation 3.29 Mean Absolute Error (MAE)*

$$MSE = \frac{1}{n} \sum_{i=1}^{n} (Y_t - F_t)^2$$

*Equation 3.30 Mean Squared Error (MSE)*

$$RMSE = \sqrt{\frac{1}{n} \sum_{i=1}^{n} (Y_t - F_t)^2}$$

*Equation 3.31 Root-Mean-Square Error (RMSE)*

Where $Y_t$ are the current values of the series and $F_t$ are the predicted values of the series

$$MASE = \frac{1}{n} \sum_{t=1}^{n} |q(t)|$$

*Equation 3.32 Mean Absolute Scaled Error (MASE)*

Where q(t) is calculated the following way

$$q(t) = \frac{V(t) - P(t)}{\sum_{i=2}^{n} \frac{V(i) - V(i-1)}{n-1}}$$

*Equation 3.33*

Where:

*V(t)*= Historical Value in period t

*P(t)*= The value of the ex-post forecast in period t

*n*: The number of periods for which the ex-post forecast is calculated

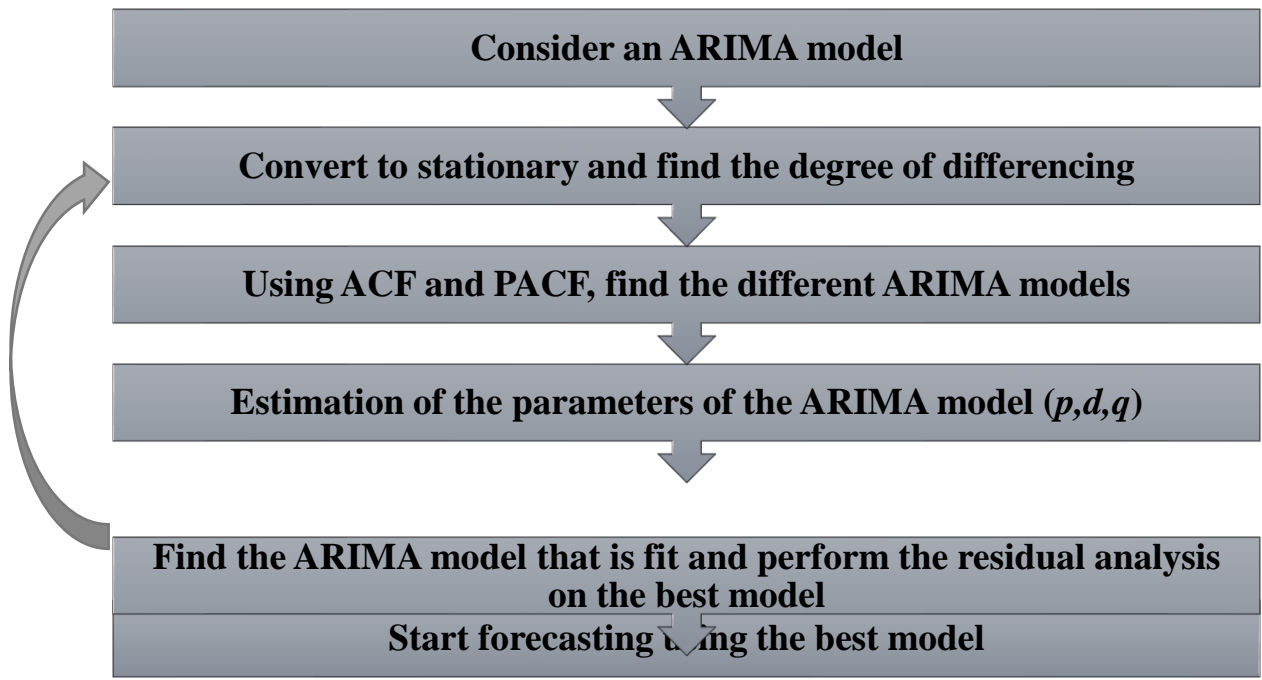

*Figure 3.1 Flow Chart of ARIMA model*

# CHAPTER FOUR

## 4 DATA ANALYSIS, RESULTS AND DISCUSSION

### 4.1 Introduction

The time series data (Q1 2013 to Q3 2022) were used to study the trend pattern of Insurance penetration rate in Ghana, to select the best ARIMA model by using various model criterion methods and to forecast future trend of penetration rate for the next thirteen quarters. The figures below indicate a time series plot of insurance penetration rate in Ghana from the first quarter of 2013 to the third quarter of 2022. The membership registration data were classified life, non-life and total.

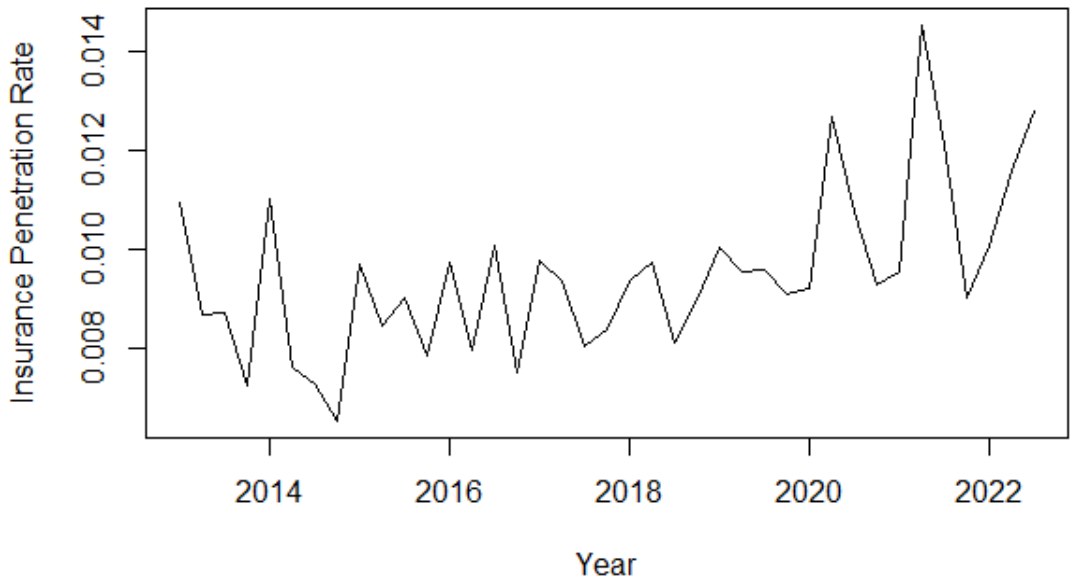

*Figure 4.1 A time series plot of insurance penetration rate in Ghana*

The time series plot (*Fig. 4.1*) of insurance penetration showed that the series was not stationary. The data was differenced to make it stationary. The first difference was enough to make the data stationary. As shown in *Fig. 4.2*, the first differenced series fluctuates about the zero-point indicating constant mean and variance, which affirms that the series is stationary.

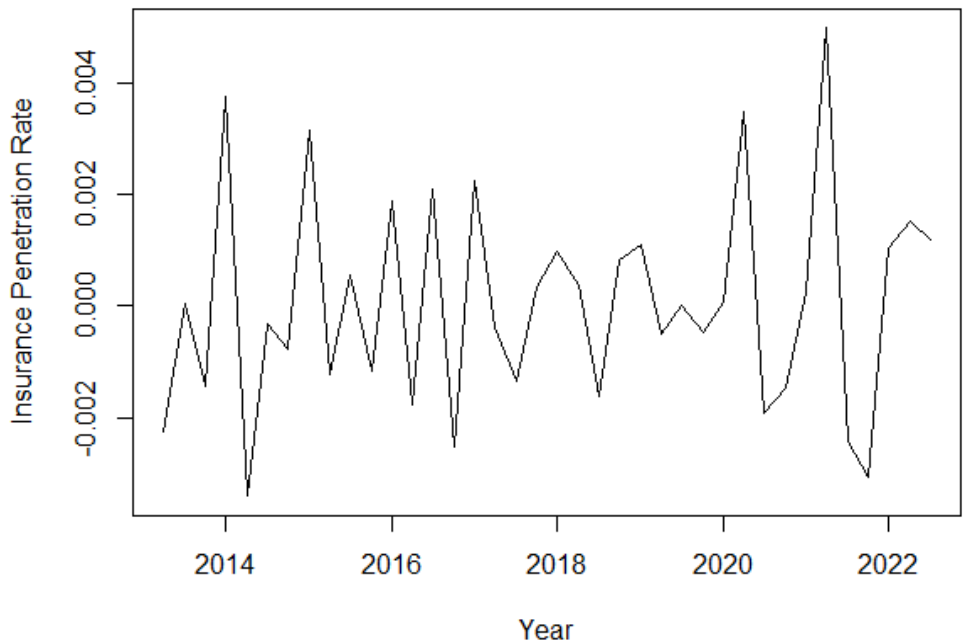

*Figure 4.2 The first differenced time series plot of insurance penetration rate in Ghana*

## 4.2 Summary Statistics for Insurance Penetration Rate

*Table 4.1 Summary statistics For Insurance Penetration rate*

| Minimum | 1st Quartile | Median | Mean | 3rd Quartile | Maximum |
|---|---|---|---|---|---|
| 0.006535 | 0.008437 | 0.009373 | 0.009491 | 0.010059 | 0.014532 |

## 4.3  The Sample Autocorrelation Function (ACF) and Partial Autocorrelation Function (PACF)

Autocorrelation is a statistical method used to measure the relationship between observations of a time series at different points in time. The autocorrelation function (ACF) measures the correlation between a time series and a lagged version of itself. The ACF at lag k is the correlation coefficient between the original time series and the time series shifted by k time units.

The ACF table and diagram provides information on the strength and direction of the correlation between the series and a lagged version of itself, which can be useful in identifying patterns and relationships in the time series data.

*Table 4.2 Autocorrelations of series 'IPRtseries', by lag*

| Lag | ACF |
|---|---|
| 0 | 1.000 |
| 1 | 0.264 |
| 2 | 0.111 |
| 3 | 0.143 |
| 4 | 0.649 |
| 5 | 0.289 |
| 6 | 0.047 |
| 7 | -0.016 |
| 8 | 0.254 |
| 9 | 0.156 |

| 10 | 0.029 |
| --- | --- |
| 11 | 0.157 |
| 12 | 0.107 |
| 13 | 0.070 |
| 14 | -0.094 |
| 15 | -0.130 |

Apart from lag zero = 4 in the ACF plot, which is expected, all the lags lie within the blue dotted lines which are the significant bounds and thus the autocorrelation for sample forecast do not exceed the significant bounds for lags between 0 and 15

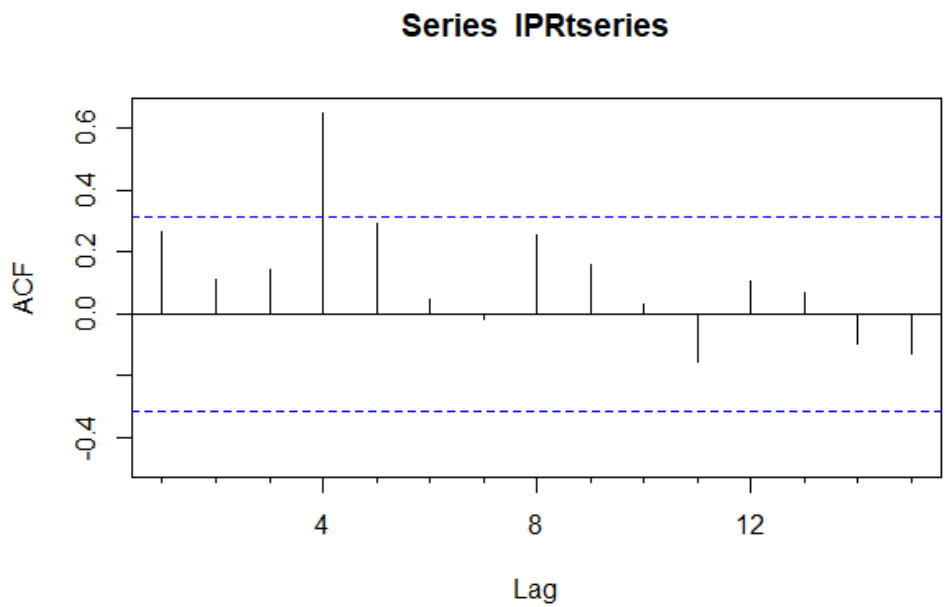

*Figure 4.3 ACF of the insurance penetration rate*

The Partial autocorrelation function (PACF) is a statistical method used to measure the correlation between observations of a time series at different points in time while controlling for the effects of intermediate lags. PACF is a useful tool in time series analysis for identifying the number of autoregressive terms in a time series model.

The PACF table and diagram provides information on the strength and direction of the correlation between the series and a lagged version of itself while controlling for the effects of intermediate lags. This can be useful in identifying the number of autoregressive terms to include in a time series model.

*Table 4.3 Partial autocorrelations of series 'IPRtseries', by lag*

| Lag | PACF |
| --- | --- |
| 1 | 0.264 |
| 2 | 0.045 |
| 3 | 0.111 |
| 4 | 0.635 |
| 5 | 0.020 |
| 6 | -0.134 |
| 7 | -0.144 |
| 8 | -0.234 |
| 9 | -0.104 |
| 10 | 0.085 |
| 11 | -0.093 |
| 12 | 0.229 |

| 13 | 0.070  |
| 14 | -0.284 |
| 15 | 0.150  |

*Figure 4.4 PACF of stationary series of insurance penetration rate*

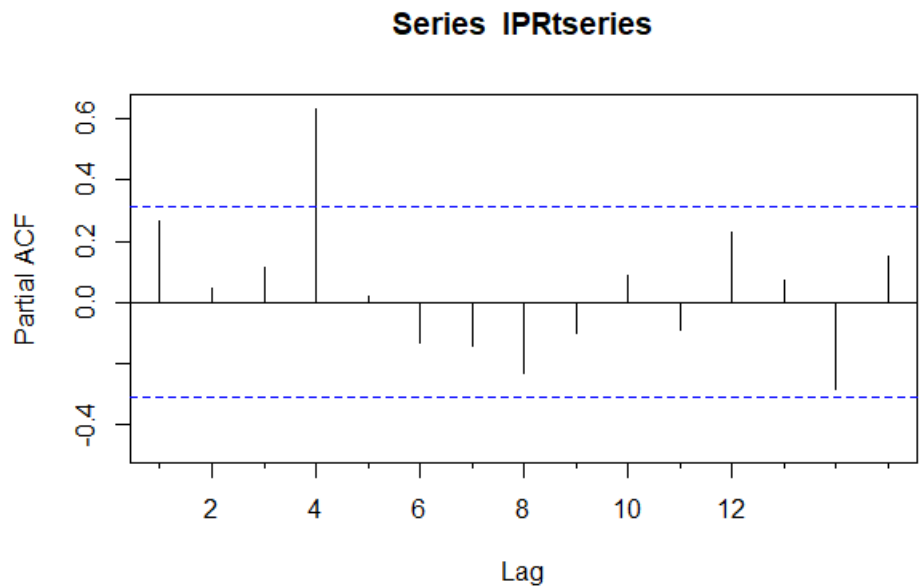

## 4.4 Test for Stationarity

### 4.4.1 Augmented Dickey-Fuller Test

Augmented Dickey Fuller test (ADF Test) is a common statistical test used to test whether a given Time series is stationary or not. It is one of the most commonly used statistical test when it comes to analysing the stationary of a series. The ADF test belongs to a category of tests called 'Unit Root Test', which is the proper method for testing the stationarity of a time series. Unit root is a characteristic of a time series that makes it non-stationary.

Since the null hypothesis assumes the presence of unit root, the p-value obtained should be less than the significance level 0.05 in order to reject the null hypothesis. Thereby, inferring that the series is stationary.

From table 4.4, the p-value is greater than the significance level 0.05, therefore we accept the null hypothesis that the insurance penetration time series data has a unit root and is not stationary.

Data: IPRtseries

*Table 4.4 Augmented Dickey-Fuller Test*

| Dickey-Fuller | Lag order | p-value |
|---|---|---|
| -2.4717 | 3 | 0.3887 |

### 4.4.2 KPSS Test for Level Stationarity

Kwiatkowski-Phillips-Schmidt-Shin (KPSS) test is a statistical test to check for stationarity of a series around a deterministic trend. Like ADF test, the KPSS test is also commonly used to analyse the stationarity of a series. A key difference from ADF test is the null hypothesis of the KPSS test is that the series is stationary. So practically, the interpretation of p-value is just the opposite to each other. From table 4.4 the p-value is less than the significance level 0.05, therefore we reject the null hypothesis that the insurance penetration time series data does not have a unit root and is stationary. Thereby accepting the alternative hypothesis that the series is not stationary.

Data: IPRtseries

*Table 4.5 KPSS Test for Level Stationarity*

| KPSS Level | Truncation lag parameter | p-value |
|---|---|---|
| 0.81768 | 3 | 0.01 |

## 4.5 Identification and Selection of Tentative ARIMA Models

The models were identified and selected by taking several combinations of orders of non-seasonal AR term and MA term *(p, d, q)* and seasonal AR and MA term *(P, D, Q)*. After finding the values of AIC and BIC, from different combination as described in table 4.6 below

*Table 4.6 Model Statistics for tentative ARIMA models*

| Model | AIC |
|---|---|
| ARIMA(2,1,2)(1,0,1)[4] with drift | Inf |
| ARIMA(0,1,0) with drift | -363.4636 |
| ARIMA(1,1,0)(1,0,0)[4] with drift | -385.3984 |
| ARIMA(0,1,1)(0,0,1)[4] with drift | Inf |
| ARIMA(0,1,0) | -365.4395 |
| ARIMA(1,1,0) with drift | -367.6584 |
| ARIMA(1,1,0)(2,0,0)[4] with drift | -384.1258 |
| ARIMA(1,1,0)(1,0,1)[4] with drift | -387.4249 |
| ARIMA(1,1,0)(0,0,1)[4] with drift | Inf |
| ARIMA(1,1,0)(2,0,1)[4] with drift | -385.8372 |
| ARIMA(1,1,0)(1,0,2)[4] with drift | -386.2225 |
| ARIMA(1,1,0)(0,0,2)[4] with drift | -387.1211 |
| ARIMA(1,1,0)(2,0,2)[4] with drift | -384.2967 |
| ARIMA(0,1,0)(1,0,1)[4] with drift | Inf |
| ARIMA(2,1,0)(1,0,1)[4] with drift | -390.1843 |
| ARIMA(2,1,0)(0,0,1)[4] with drift | Inf |
| ARIMA(2,1,0)(1,0,0)[4] with drift | -388.7587 |
| ARIMA(2,1,0)(2,0,1)[4] with drift | -388.6882 |
| ARIMA(2,1,0)(1,0,2)[4] with drift | -388.8122 |
| ARIMA(2,1,0)          with drift | -369.6004 |
| ARIMA(2,1,0)(0,0,2)[4] with drift | -389.7061 |
| ARIMA(2,1,0)(2,0,0)[4] with drift | -387.3178 |
| ARIMA(2,1,0)(2,0,2)[4] with drift | -386.9079 |
| ARIMA(3,1,0)(1,0,1)[4] with drift | -396.2369 |
| ARIMA(3,1,0)(0,0,1)[4] with drift | -395.6926 |
| ARIMA(3,1,0)(1,0,0)[4] with drift | -395.5609 |
| ARIMA(3,1,0)(2,0,1)[4] with drift | Inf |
| ARIMA(3,1,0)(1,0,2)[4] with drift | Inf |
| ARIMA(3,1,0)          with drift | -397.5471 |
| ARIMA(3,1,1)          with drift | -396.2451 |
| ARIMA(2,1,1)          with drift | -381.2976 |
| ARIMA(3,1,0) | -398.502 |
| ARIMA(3,1,0)(1,0,0)[4] | -396.6313 |
| ARIMA(3,1,0)(0,0,1)[4] | -397.477 |
| ARIMA(3,1,0)(1,0,1)[4] | -397.8927 |
| ARIMA(2,1,0) | -371.4608 |
| ARIMA(3,1,1) | -396.8921 |
| ARIMA(2,1,1) | -380.8745 |

The distribution of variables in Fig.4.1 and the tests performed for stationarity in Table 4.4 and Table 4.5 shows that the data is non-stationary. The ARIMA technique is used to describe the necessary degree of differencing and when applied makes the data stationary. The auto.arima function in Rstudio derived the optimal ARIMA model for IPR as well as the lowest Akaike Information Criteria (AIC) which gives the best model.

The software RStudio (The expert modeller), automatically confirms the best-fitting model for each dependent series. Through expert modeller, the model variables have been transformed where appropriate, using differencing and/or a square root or natural log transformation. The suggested model obtained by RStudio is Non-Seasonal ARIMA (3, 1, 0)

## 4.6 Model Analysis

Series: Insurance Penetration Rate Time Series

ARIMA(*3,1,0*)

*Table 4.7 Autoregressive Coefficients*

| | Coefficients | | |
|---|---|---|---|
| | ar1 | ar2 | ar3 |
| | -0.8305 | -0.7782 | -0.7592 |
| s.e | 0.1100 | 0.1173 | 0.1020 |

*Equation 4.1 Best ARIMA Model*

$$Y'_t = -0.8305 Y'_{t-1} - 0.7782 Y'_{t-2} - 0.7592 Y'_{t-3} + \varepsilon_t$$

$$\text{Where } Y'_t = Y_t - Y_{t-1}$$

Also where $\varepsilon_t$ is white noise with standard deviation $\sqrt{1.327 \times 10^{-6}} = 1.152 \times 10^{-3}$

*Table 4.8 Model Estimation*

| sigma^2($\sigma^2$) | log likelihood | AIC | AICc | BIC |
|---|---|---|---|---|
| 1.327x10$^{-6}$ | 203.25 | -398.5 | -397.29 | -391.95 |

| | ME | RMSE | MAE | MPE | MAPE | MASE | ACF1 |
|---|---|---|---|---|---|---|---|
| Training set error measures | | | | | | | |
| Training set | 0.0001097798 | 0.001091171 | 0.0008645403 | 0.01008047% | 9.131175% | 0.9477387 | -0.0580795 |

The sigma^2($\sigma^{2)}$ is the variance of the error terms thus, $\varepsilon$ (t). The variance of $\varepsilon(t)$ is $1.327 \times 10^{-6}$.

Log likelihood is the value of measure of goodness of fit for any model.

The Akaike Information Criterion (AIC) is a function of the log likelihood. AIC is used to compare different models and determine which one is best for the data. The model with the least AIC is selected. AICc is the sample-size corrected Akaike Information Criterion

Bayesian Information Criterion is based in part of the likelihood function and is closely related to AIC. It is also used to determine which model is best.

The Mean Error (ME) is the average of all errors in a set. It is biased due to the offsetting effect of positive and negative forecast errors. The ME can quickly represent the symmetry of the error distribution, which can be useful in assessing a specific model. A positive value means that the predicted value is less than the true value. Our ME of 0.0001098 is negligible compared to most of the measured values.

From table 4.8, the Root Mean Square Error (RMSE) is 0.001091, the closer the RMSE is to 0, the more accurate the model is. Simply put, RMSE can be interpreted as the average error that the model's predictions have in comparison with the actual, whereby extra weight is added to larger prediction error by finding the square root of the MSE. Therefore, an RMSE of 0.001091 is a good measure of accuracy.

Mean Absolute Error (MAE) can be interpreted as the average error that the model's predictions have in comparison with their corresponding observed values. However, MAE is returned on the same scale as the target you are predicting for. The closer the MAE is to zero, the more accurate the model. Hence, an MAE of 0.0008645 is a good measure of accuracy.

Mean Percentage Error (MPE) is the mean percentage error (or deviation). It is a relative measure that essentially scales ME to be in percentage units instead of the variable's units. Unlike MAE and MAPE, MPE is useful because it allows us to see if our model systematically underestimates (error that is more negative or over overestimates (positive error). In this case, the MPE is 0.01008% indicating a positive error.

Mean Absolute Percentage Error (MAPE), is the percentage equivalent of MAE but with adjustments to convert everything into percentages for people to conceptualize easily.

| MAPE | Interpretation |
| --- | --- |
| <10% | Very good |
| 10% - 20% | Good |
| 20%-50% | Ok |
| >50% | Not Good |

The model has an MAPE of 9.13% hence a very good measure of accuracy.

The mean absolute scaled error (MASE) of one data point is defined as the (one-period-ahead) forecast error divided by the average forecast error of the naïve method. The MASE can be used to compare forecast methods on a single time series and to compare forecast accuracy between series. MASE was proposed by Koehler & Hyndman (2006).

If MASE < 1, the one it arises from a better forecast than the average one-step, naïve forecast computed in sample.

IF MASE > 1, the forecast is worse than the average one-step, naïve forecast computed in sample.

Since the MASE for this model is 0.9477, which is less than one, which implies that, the actual forecast. A value of 0.9477 is close to one. Which means that our model is exactly good as just picking the last observation. Nevertheless, a lower value, the better.

Autocorrelation of errors at lag 1(ACF1) is a measure of how much the current value is influenced by the previous values in time series. ACF1 of -0.0580795 means the correlation between a point and the next point is -0.0580795.

## 4.7    Diagnostic Checking

### 4.7.1    Residual Diagnostic Testing Using the Ljung-Box Test

The residuals for the fitted models were computed and a diagnostic test using the Ljung-Box test was used to further confirm that the residuals have no autocorrelation and the models fit the data very well.

Data:  Residuals from ARIMA(3,1,0)

*Table 4.9 Ljung-Box Test on Residuals*

| Q* | df | p-value | Model df | Total lags used |
|---|---|---|---|---|
| 5.1367 | 5 | 0.3994 | 3 | 8 |

From Table 4.9 since the p-value is greater than 0.05 we cannot reject the null hypothesis hence we confirm that our models are a good fit to forecast IPR.

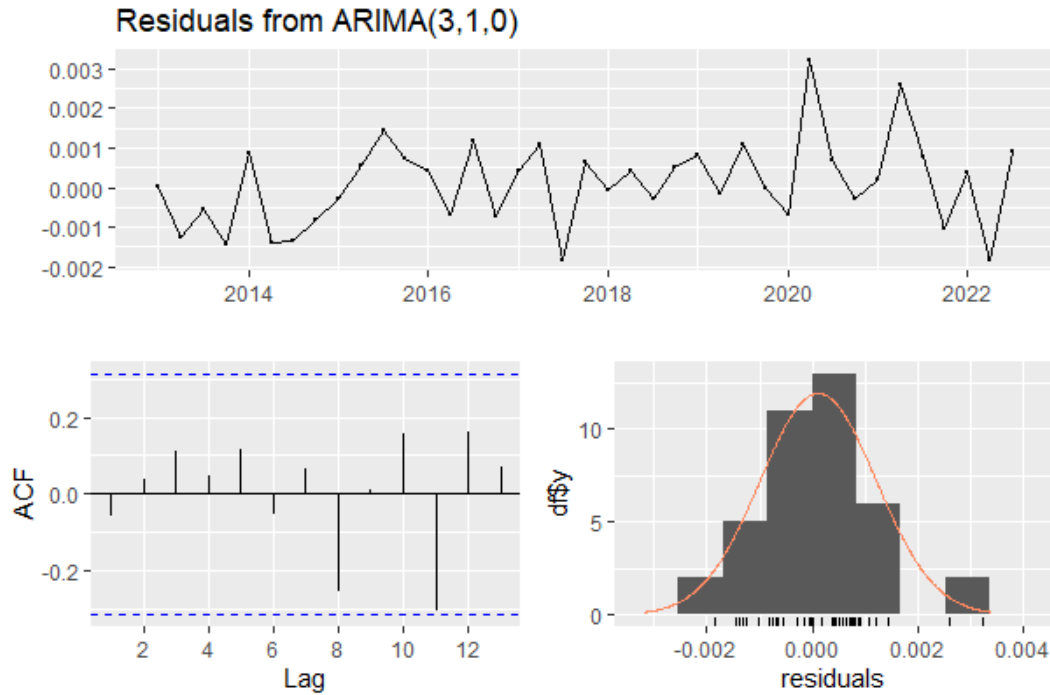

*Figure 4.5 Residuals from ARIMA (3,1,0)*

### 4.7.2   Test for Stationarity of the Model Residuals

Our null hypothesis ($H_0$) in the test is that the model residuals data is non-stationary while alternative hypothesis ($H_1$) is that the series is stationary. The hypothesis then is tested by applying the Augmented Dickey-Fuller Test (ADF) and Phillips-Perron Unit Root Test (PP) to the model residuals data. The ADF and PP test result, as obtained upon application, is shown below:

### 4.7.3   Augmented Dickey-Fuller Test

From table 4.10 the p-value is less than the significance level 0.05, therefore we reject the null hypothesis that the insurance penetration time series data has a unit root and is not stationary. The alternative hypothesis is accepted.

Data:  IPRmodel$residuals

*Table 4.10 Augmented Dickey-Fuller Test*

| Dickey-Fuller | Lag order | p-value |
|---|---|---|
| -4.5236 | 1 | 0.01 |

### 4.7.4 Phillips-Perron Unit Root Test

The Phillips-Perron test is similar to the ADF test, but it is a bit more advanced. It checks to see if the data points are changing in a predictable way. If the data points are changing in a predictable way, then the time series is stationary. If the p-value is above the significance level of 0.05, then the null hypothesis cannot be rejected.

From table 4.11 the p-value is less than the significance level 0.05, therefore we reject the null hypothesis that the insurance penetration time series data has a unit root and is not stationary. The alternative hypothesis is accepted.

Data: IPRmodel$residuals

*Table 4.11 Phillips-Perron Unit Root Test*

| Dickey-Fuller Z(alpha) | Truncation lag parameter | p-value |
|---|---|---|
| -42.431 | 3 | 0.01 |

We, therefore, fail to accept the $H_0$ and hence can conclude that the alternative hypothesis is true i.e. the series is stationary in its mean and variance. Thus, there is no need for further differencing the time series and we adopt $d = 1$ for our ARIMA($p,d,q$) model.

### 4.7.5 KPSS Test for Level Stationarity

From table 4.12 the p-value is greater than the significance level 0.05, therefore we accept the null hypothesis that the insurance penetration time series data does not have a unit root and is stationary.

Data: IPRmodel$residuals

*Table 4.12 KPSS Test for Level Station*

| KPSS Level | Truncation lag parameter | p-value |
| --- | --- | --- |
| 0.37599 | 3 | 0.0875 |

### 4.7.6 ACF and PACF of The Model Residuals

#### *4.7.6.1 ACF of residuals*

In the table and diagram provided below, the ACF values are shown for different lags (or time intervals) of the time series. The ACF values range from -1 to 1, where a value of 1 indicates a perfect positive correlation, 0 indicates no correlation, and -1 indicates a perfect negative correlation. At lag 0, the ACF is always 1 since the correlation of a series with itself at the same time is always 1. As for the other lags, the values indicate the strength and direction of the correlation between the series and a lagged version of itself.

For instance, at lag 1, the ACF value is -0.058, indicating a weak negative correlation between the series and the lagged version of itself by one time unit. At lag 2, the ACF value is 0.036, indicating a weak positive correlation between the series and the lagged version of itself by two time units. Similarly, at lag 3, the ACF value is 0.109, indicating a moderate positive correlation between the series and the lagged version of itself by three time units. At lag 8, the ACF value is -0.254, indicating a moderate negative correlation between the series and the lagged version of itself by eight time units.

*Table 4.13 Autocorrelations of series 'ts(IPRmodel$residuals)', by lag*

| Lag | ACF |
|---|---|
| 0 | 1.000 |
| 1 | -0.058 |
| 2 | 0.036 |
| 3 | 0.109 |
| 4 | 0.045 |
| 5 | 0.116 |
| 6 | -0.055 |
| 7 | 0.064 |
| 8 | -0.254 |
| 9 | 0.011 |
| 10 | 0.155 |
| 11 | -0.304 |
| 12 | 0.159 |
| 13 | 0.070 |
| 14 | -0.106 |
| 15 | 0.072 |

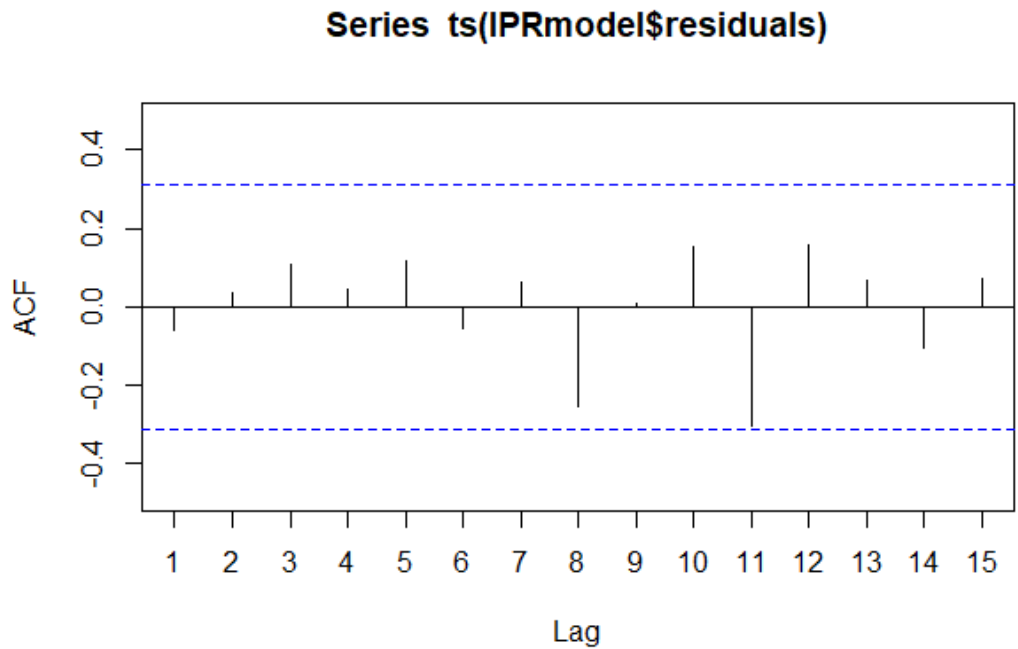

*Figure 4.6 ACF Correlogram of residuals*

### *4.7.6.2 PACF of residuals*

In the table and diagram provided, the PACF values are shown for different lags (or time intervals) of the time series. The PACF values range from -1 to 1, where a value of 1 indicates a perfect positive correlation, 0 indicates no correlation, and -1 indicates a perfect negative correlation.

At lag 1, the PACF value is -0.058, indicating a weak negative correlation between the series and the lagged version of itself by one time unit while controlling for the effects of intermediate lags.

At lag 2, the PACF value is 0.033, indicating a weak positive correlation between the series and the lagged version of itself by two time units while controlling for the effects of intermediate lags.

Similarly, at lag 3, the PACF value is 0.114, indicating a moderate positive correlation between the series and the lagged version of itself by three time units while controlling for the effects of intermediate lags. At lag 8, the PACF value is -0.284, indicating a moderate negative correlation between the series and the lagged version of itself by eight time units while controlling for the effects of intermediate lags.

*Table 4.14 Partial autocorrelations of series 'ts(IPRmodel$residuals)', by lag*

| Lag | PACF |
| --- | --- |
| 1 | -0.058 |
| 2 | 0.033 |
| 3 | 0.114 |
| 4 | 0.058 |
| 5 | 0.116 |
| 6 | -0.058 |
| 7 | 0.039 |
| 8 | -0.284 |
| 9 | -0.026 |
| 10 | 0.164 |
| 11 | -0.245 |
| 12 | 0.186 |

| 13 | 0.158 |
| 14 | -0.154 |
| 15 | 0.094 |

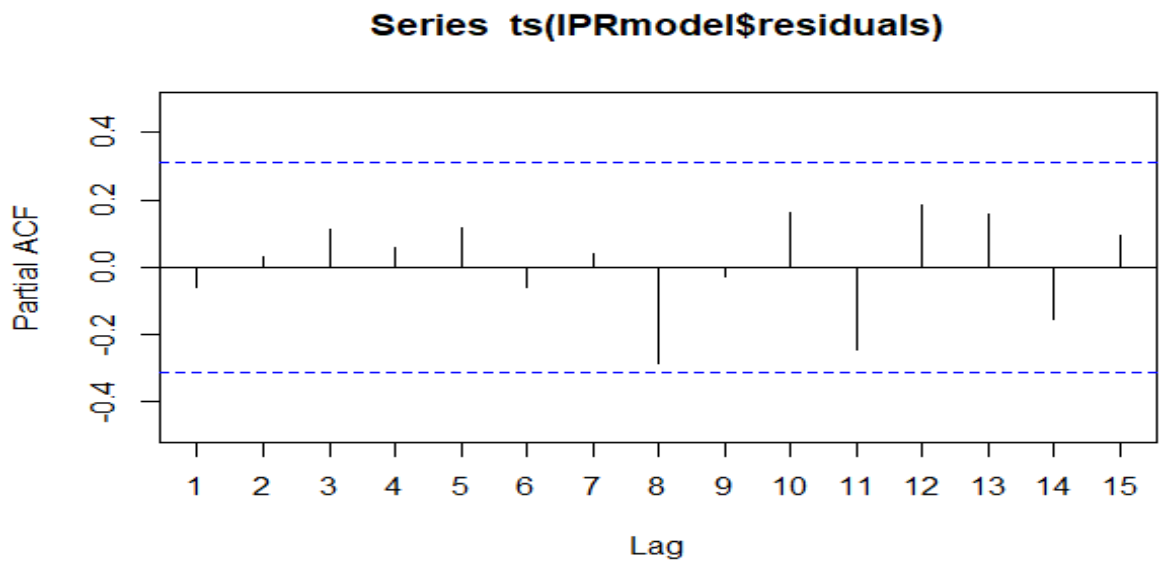

*Figure 4.7 PACF Correlogram of Residual*

## 4.8 Forecasting

### 4.8.1 Insurance Penetration Forecast at 95% Confidence

*Table 4.15 Forecasts of Quarterly Insurance Penetration Rate from 2022-2024 at 95% confidence*

| Point | Forecast | Low at 95% | High at 95% |
| --- | --- | --- | --- |
| 2022 Q4 | 0.009813194 | 0.007555635 | 0.01207075 |
| 2023 Q1 | 0.010188966 | 0.007899211 | 0.01247872 |
| 2023 Q2 | 0.011285177 | 0.008988131 | 0.01358222 |
| 2023 Q3 | 0.012331074 | 0.010032110 | 0.01463004 |
| 2023 Q4 | 0.010324032 | 0.007436748 | 0.01321132 |
| 2024 Q1 | 0.010344669 | 0.007396726 | 0.01329261 |
| 2024 Q2 | 0.011095404 | 0.008128740 | 0.01406207 |
| 2024 Q3 | 0.011979663 | 0.009006793 | 0.01495253 |
| 2024 Q4 | 0.010645356 | 0.007364662 | 0.01392605 |
| 2025 Q1 | 0.010495367 | 0.007140154 | 0.01385058 |
| 2025 Q2 | 0.010986983 | 0.007601991 | 0.01437198 |
| 2025 Q3 | 0.011708469 | 0.008311515 | 0.01510542 |
| 2025 Q4 | 0.010840548 | 0.007257901 | 0.01442320 |

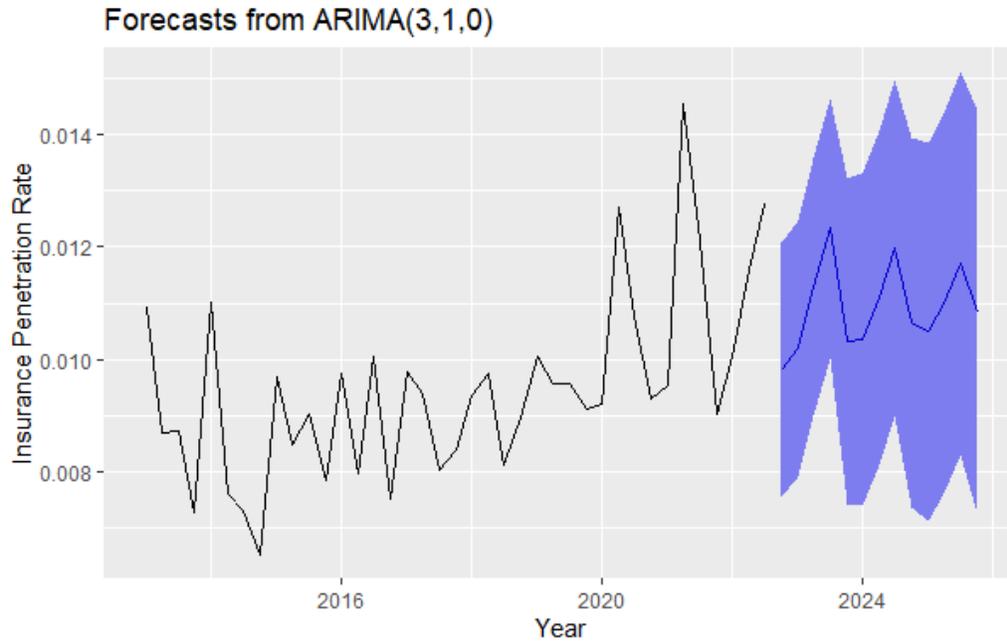

*Figure 4.8 Forecasts of Quarterly Insurance Penetration Rate from 2022-2024 at 95% confidence.*

### 4.8.1.1 *Box-Ljung test*
Data: IPRforecast

*Table 4.16 Box-Ljung test of forecasted insurance penetration rates (95% confidence)*

| X-squared | df | p-value |
|---|---|---|
| 1.4557 | 5 | 0.9181 |
| 6.4672 | 10 | 0.7746 |

### 4.8.2 Insurance Penetration Forecast at 99% Confidence

*Table 4.17 Forecasts of Quarterly Insurance Penetration Rate from 2022-2024 at 99% confidence*

| Point   | Forecast    | Low at 95%  | High at 95% |
|---------|-------------|-------------|-------------|
| 2022 Q4 | 0.009813194 | 0.006846258 | 0.01278013  |
| 2023 Q1 | 0.010188966 | 0.007179718 | 0.01319821  |
| 2023 Q2 | 0.011285177 | 0.008266347 | 0.01430401  |
| 2023 Q3 | 0.012331074 | 0.009309723 | 0.01535242  |
| 2023 Q4 | 0.010324032 | 0.006529497 | 0.01411857  |
| 2024 Q1 | 0.010344669 | 0.006470416 | 0.01421892  |
| 2024 Q2 | 0.011095404 | 0.007196546 | 0.01499426  |
| 2024 Q3 | 0.011979663 | 0.008072650 | 0.01588668  |
| 2024 Q4 | 0.010645356 | 0.006333793 | 0.01495692  |
| 2025 Q1 | 0.010495367 | 0.006085869 | 0.01490487  |
| 2025 Q2 | 0.010986983 | 0.006538349 | 0.01543562  |
| 2025 Q3 | 0.011708469 | 0.007244114 | 0.01617282  |
| 2025 Q4 | 0.010840548 | 0.006132151 | 0.01554895  |

*Figure 4.9 Forecasts of Quarterly Insurance Penetration Rate from 2022-2024 at 99% confidence*

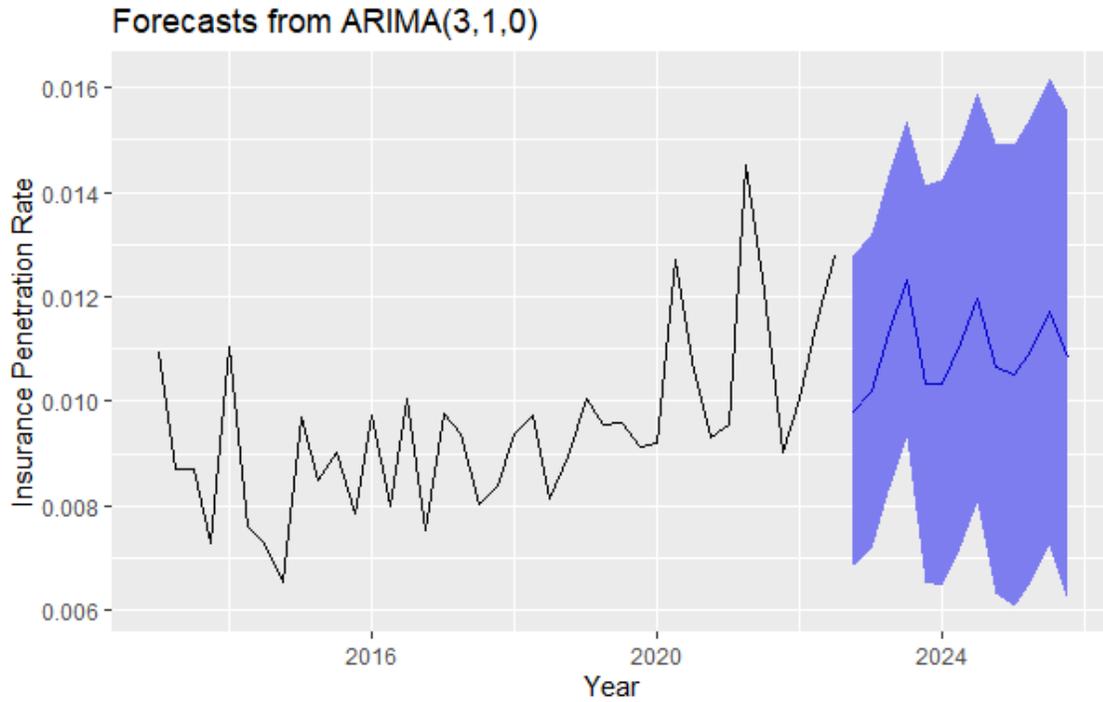

### 4.8.2.1 *Box-Ljung test*
Data: IPRforecast

*Table 4.18 Box-Ljung test of forecasted insurance penetration rates (99% confidence)*

| X-squared | df | p-value |
|---|---|---|
| 1.4557 | 5 | 0.9181 |

# CHAPTER FIVE

## 5. SUMMARY, CONCLUSIONS & RECOMMENDATIONS

### 5.1 Summary

Insurance penetration is one of the main determinants of a country's financial development. Forecasting helps stakeholders anticipate future events and provide fact-based support to the decision-making process. In this context, ARIMA model is used to predict insurance penetration in Ghana. Post-estimation validation shows that the residuals ACF and PACF were within the SE limits. The Mean Absolute Percentage Error (MAPE) is about 9.13%, which means the model is about 90.87% accurate. This rate highlights ARIMA's reliability in predicting future values.

### 5.2 Conclusion

A key finding of this research is the increase in insurance penetration in Ghana but at low staggering rate. Low insurance penetration results in weaker economic resilience of the uninsured. However, the results may alert stakeholders involved to correct, hasten or intensify the speed of increasing penetration rate to match Namibia's 6.69% and eventually South Africa's 17%.

Insurance is a vital component of financial services that provides protection against unforeseen events, losses, and damages. In Ghana, insurance penetration remains low, with many people not having insurance coverage. According to the Ghana Insurance Report (2019), insurance penetration was at 1.1% in 2018, indicating that only a small portion of the population has insurance coverage. The low insurance penetration rate can be attributed to several factors, including low awareness, poor perception of insurance and inadequate marketing strategies.

## 5.3 Recommendations

Therefore, this study will also explore various strategies that can be employed to increase insurance penetration rate in Ghana.

Insurance can be improved through:

**Bancassurance**

Bancassurance is a relationship between a bank and an insurance company that is aimed at offering insurance products or insurance benefits to the bank's customers.

Banking institutions and insurance companies have found bancassurance to be an attractive and profitable addition to their existing operations. Ghana's insurance industry relies heavily on agents and brokers to sell insurance products. Intermediation and broker-driven channels have not achieved significant insurance penetration. Insurance industry players must find new and more efficient ways. Bancassurance is recognized as one of the distribution channels that can improve insurance penetration.

Developing annuity markets to enable the life insurance sector to optimize synergies with the pensions sector

**Improving Public Awareness**

Low public awareness is one of the primary reasons for the low insurance penetration rate in Ghana. Many people are unaware of the benefits of insurance and how it works. Insurance companies must invest in public awareness campaigns to educate Ghanaians about insurance and the protection it offers. These campaigns can be in the form of advertisements on TV, radio, and billboards. Insurance companies can also engage in community outreach programs to educate people about insurance.

**Developing Tailored Products**

One of the reasons why people do not have insurance coverage in Ghana is that they cannot afford the premiums. Insurance companies need to develop tailored products that meet the needs of the people. These products should be affordable and offer adequate protection. For example, an insurance company can develop a health insurance product that covers only basic medical expenses, which is affordable for people with low income.

**Enhancing Product Distribution**

Another challenge facing insurance companies in Ghana is distribution. Many people do not have access to insurance products because they do not know where to find them. Insurance companies need to expand their distribution channels to reach more people. They can collaborate with banks and mobile network operators to offer insurance products through their platforms. This will make it easier for people to access insurance products.

**Enhancing Customer Experience**

Insurance companies need to improve the customer experience to increase retention rates. They can achieve this by investing in technology to automate the claims process and reduce waiting times. Insurance companies can also offer personalized services to their customers, such as customized insurance policies and dedicated account managers.

**Developing Strategic Partnerships**

Insurance companies can also increase insurance penetration rates in Ghana by developing strategic partnerships with other organizations. For example, an insurance company can collaborate with a hospital to offer health insurance products to its patients. Insurance companies can also collaborate with schools to offer insurance products to students.

**Providing Incentives**

Insurance companies can also provide incentives to encourage people to purchase insurance products. For example, an insurance company can offer discounts to customers who purchase multiple insurance products. They can also offer rewards for customers who have not made any claims in a year.

**Educating Sales Agents**

Insurance companies can also increase insurance penetration rates by educating their sales agents. Sales agents should be knowledgeable about insurance products and how they work. They should be able to explain the benefits of insurance to potential customers and answer their questions. Insurance companies can invest in training programs for their sales agents to improve their knowledge and skills.

**Using Social Media**

Social media is an effective platform for reaching a large audience. Insurance companies can use social media platforms such as Facebook, Twitter, and Instagram to promote their products and services. They can also use social media to engage with their customers and address their concerns.

**Offering Online Services:**

Insurance companies can also offer online services to increase convenience for their customers. Customers can purchase insurance products online and manage their policies through an online portal. This will make it easier for people to access insurance products and services.

**Government Support**

Finally, the Ghanaian government can play a role in increasing insurance penetration rates by providing support to the insurance industry. The government can offer tax incentives to insurance companies to encourage them to expand their operations.

## 5.4 Limitations of the Study

Other jurisdictions, such as Kenya and South Africa, consider health insurance and pensions when determining penetration. Assuming that Ghana's penetration measure includes both health insurance and pensions, the estimated penetration is about 3%.

The most conventional tool used to gauge the development of a country's insurance market is the insurance penetration rate. Yet while it serves as a broad, high-level indicator of the insurance market's development, the penetration rate does not reveal detailed information about the actual dynamics of the local insurance market. It does not indicate how many people actually have insurance coverage, nor does it signify the quality of coverage and whether it provides value to clients. For supervisors who have enhancing access to insurance as part of their mandate and/ or want to get a better picture of client value, the insurance penetration rate is unlikely to be sufficiently meaningful as only limited information can be drawn from it.

There are currently 27 microinsurance products in Ghana, majority of which are Mobile Network Operators (MNOs). Spanning a vast variety of risks and covering 7.5 million lives, or 28% of the population. The majority of these individuals are low-income earners in the informal sector who work under precarious conditions and who would otherwise not have access to insurance. The value that can be derived from such policies is greatly significant and is not reflected in the penetration rate or density figures. The rapid growth of microinsurance in Ghana has been largely impacted by the prevalence of mobile microinsurance products.

## 5.5    Suggestions for Further Research

Future research should examine the reasons behind the stagnancy in Ghana's insurance penetration. It is worth using different insurance penetration proxies and forecasting techniques to check the robustness of the results. Other forecasting techniques such as Holt-Winters Exponential Smoothing and Artificial Neural Networks (ANNs).

# APPENDIX

## Appendix 1: Raw Data

Source:

Ghana Statistical Service (Quarterly Gross Domestic Product)

National Insurance Commission (Quarterly Gross Premiums)

| Year Quarter | Life Insurance Gross Premium (GHS) | Non-Life Insurance Gross Premium (GHS) | Total Gross Premium (GHS) | Quarterly GDP (GHS) | Life Insurance Penetration Rate | Non-Life Insurance Penetration Rate | Total Insurance Penetration Rate |
|---|---|---|---|---|---|---|---|
| 2013_Q1 | 103,575,423 | 203,013,013 | 306,588,436 | 28038900000 | 0.3694% | 0.7240% | 1.0934% |
| 2013_Q2 | 111,557,839 | 137,858,957 | 249,416,796 | 28715700000 | 0.3885% | 0.4801% | 0.8686% |
| 2013_Q3 | 121,279,872 | 130,855,909 | 252,135,781 | 28924800000 | 0.4193% | 0.4524% | 0.8717% |
| 2013_Q4 | 132,842,369 | 100,906,111 | 233,748,480 | 32149400000 | 0.4132% | 0.3139% | 0.7271% |
| 2014_Q1 | 128,060,078 | 228,108,375 | 356,168,453 | 32301400000 | 0.3965% | 0.7062% | 1.1026% |
| 2014_Q2 | 141,180,936 | 128,938,638 | 270,119,574 | 35475400000 | 0.3980% | 0.3635% | 0.7614% |
| 2014_Q3 | 149,509,110 | 145,452,938 | 294,962,048 | 40426800000 | 0.3698% | 0.3598% | 0.7296% |
| 2014_Q4 | 153,158,105 | 114,294,570 | 267,452,675 | 40924600000 | 0.3742% | 0.2793% | 0.6535% |
| 2015_Q1 | 161,174,827 | 246,501,602 | 407,676,429 | 42016200000 | 0.3836% | 0.5867% | 0.9703% |
| 2015_Q2 | 177,487,539 | 195,416,329 | 372,903,868 | 44003500000 | 0.4033% | 0.4441% | 0.8474% |
| 2015_Q3 | 181,069,501 | 213,503,006 | 394,572,507 | 43680700000 | 0.4145% | 0.4888% | 0.9033% |
| 2015_Q4 | 188,757,037 | 197,325,426 | 386,082,463 | 49133400000 | 0.3842% | 0.4016% | 0.7858% |
| 2016_Q1 | 186,086,838 | 331,000,125 | 517,086,963 | 53070600000 | 0.3506% | 0.6237% | 0.9743% |
| 2016_Q2 | 151,433,272 | 266,791,486 | 418,224,758 | 52440400000 | 0.2888% | 0.5088% | 0.7975% |
| 2016_Q3 | 288,932,956 | 250,770,559 | 539,703,515 | 53636200000 | 0.5387% | 0.4675% | 1.0062% |
| 2016_Q4 | 225,159,770 | 222,615,869 | 447,775,639 | 59420300000 | 0.3789% | 0.3746% | 0.7536% |

| Quarter | | | | | | | |
|---|---|---|---|---|---|---|---|
| 2017_Q1 | 232,983,398 | 362,170,136 | 595,153,534 | 60905300000 | 0.3825% | 0.5946% | 0.9772% |
| 2017_Q2 | 279,683,522 | 294,209,983 | 573,893,505 | 61179200000 | 0.4572% | 0.4809% | 0.9381% |
| 2017_Q3 | 228,587,017 | 296,946,203 | 525,533,220 | 65330000000 | 0.3499% | 0.4545% | 0.8044% |
| 2017_Q4 | 320,637,820 | 236,759,996 | 557,397,816 | 66361000000 | 0.4832% | 0.3568% | 0.8399% |
| 2018_Q1 | 311,385,975 | 372,296,561 | 683,682,536 | 72938600000 | 0.4269% | 0.5104% | 0.9373% |
| 2018_Q2 | 328,299,346 | 333,330,328 | 661,629,674 | 67897100000 | 0.4835% | 0.4909% | 0.9745% |
| 2018_Q3 | 314,047,824 | 288,418,734 | 602,466,558 | 74147800000 | 0.4235% | 0.3890% | 0.8125% |
| 2018_Q4 | 381,934,294 | 305,242,741 | 687,177,035 | 76632800000 | 0.4984% | 0.3983% | 0.8967% |
| 2019_Q1 | 379,600,738 | 445,988,910 | 825,589,648 | 82086700000 | 0.4624% | 0.5433% | 1.0058% |
| 2019_Q2 | 404,317,452 | 365,496,534 | 769,813,986 | 80556800000 | 0.5019% | 0.4537% | 0.9556% |
| 2019_Q3 | 413,630,504 | 404,981,013 | 818,611,517 | 85495600000 | 0.4838% | 0.4737% | 0.9575% |
| 2019_Q4 | 451,292,249 | 342,425,436 | 793,717,684 | 87070500000 | 0.5183% | 0.3933% | 0.9116% |
| 2020_Q1 | 372,825,543 | 515,741,117 | 888,566,660 | 96590200000 | 0.3860% | 0.5339% | 0.9199% |
| 2020_Q2 | 575,397,490 | 514,426,034 | 1,089,823,524 | 85869600000 | 0.6701% | 0.5991% | 1.2692% |
| 2020_Q3 | 520,908,274 | 477,290,554 | 998,198,828 | 92699600000 | 0.5619% | 0.5149% | 1.0768% |
| 2020_Q4 | 529,659,620 | 427,280,661 | 956,940,281 | 102988500000 | 0.5143% | 0.4149% | 0.9292% |
| 2021_Q1 | 482,221,650 | 578,393,209 | 1,060,614,859 | 111105100000 | 0.4340% | 0.5206% | 0.9546% |
| 2021_Q2 | 730,339,342 | 682,541,878 | 1,412,881,220 | 97227900000 | 0.7512% | 0.7020% | 1.4532% |
| 2021_Q3 | 644,937,301 | 650,056,842 | 1,294,994,143 | 106999400000 | 0.6027% | 0.6075% | 1.2103% |
| 2021_Q4 | 655,849,926 | 450,783,749 | 1,106,633,675 | 122642800000 | 0.5348% | 0.3676% | 0.9023% |
| 2022_Q1 | 570,334,984 | 791,989,802 | 1,362,324,787 | 135425000000 | 0.4211% | 0.5848% | 1.0060% |
| 2022_Q2 | 597,580,096 | 801,636,568 | 1,399,216,665 | 120685100000 | 0.4952% | 0.6642% | 1.1594% |
| 2023_Q3 | 1,009,969,356 | 778,291,916 | 1,788,261,273 | 139980600000 | 0.7215% | 0.5560% | 1.2775% |

## Appendix 2: R Source Code

```r
# Import Dataset
class(IPR) # Check the type of Dataset
str(IPR) # The structure of the Dataset
Insurance_PR=IPR$Total_Insurance_Pentration_Rate
library(ggplot2) # Load the package ggplot2
library(ggfortify) # Load the package ggfortify
library(forecast) # Load the package forecast
library(tseries) # Load the package tseries
IPRtseries=ts(Insurance_PR, start=c(2013), frequency = 4) # convert the data variable into time series and assign a new variable name
plot.ts(IPRtseries, xlab = "Year", ylab = "Insurance Penetration Rate") # plot the time series data
Acf(IPRtseries) # plot the autocorrelation function of the time series data
print(Acf(IPRtseries))
Pacf(IPRtseries) # plot the partial autocorrelation function of the time series data
print(Pacf(IPRtseries))
adf.test(IPRtseries) # Augmented Dickey-Fuller Test determines stationarity
kpss.test(IPRtseries)
DiffIPRtseries=diff(IPRtseries)
plot.ts(DiffIPRtseries, xlab = "Year", ylab = "Insurance Penetration Rate")
adf.test(DiffIPRtseries)
IPRmodel=auto.arima(IPRtseries,ic="aic",trace = TRUE) # Determine the best ARIMA model and assign a variable name
IPRmodel # Show the coefficient of the model and the AIC and BIC
summary(IPRmodel)
checkresiduals(IPRmodel) # plots and checks the residual of the ARIMA model
adf.test(IPRmodel$residuals, k=1) # Augmented Dickey-Fuller Test determines stationary of the residuals of the model
pp.test(IPRmodel$residuals)
kpss.test(IPRmodel$residuals)
Acf(ts(IPRmodel$residuals)) # plot the autocorrelation function of the model residuals
```

```
print(Acf(ts(IPRmodel$residuals)))
Pacf(ts(IPRmodel$residuals)) # plot the partial autocorrelation function of the model residuals
print(Pacf(ts(IPRmodel$residuals)))
IPRforecast<- forecast(IPRmodel, level=c(95), h=13) # Forecast the model for the next 10
IPRforecast # Show forecasted values
autoplot(IPRforecast, xlab = "Year", ylab = "Insurance Penetration Rate") # plot forecasted values
Box.test(IPRforecast,lag = 5,type="Ljung-Box") # Apply the Ljung-Box
Box.test(IPRforecast$res,lag = 10,type="Ljung-Box") # Apply the Ljung-Box
IPRforecast<- forecast(IPRmodel, level=c(99), h=13) # Forecast the model for the next 10
IPRforecast # Show forecasted values
autoplot(IPRforecast, xlab = "Year", ylab = "Insurance Penetration Rate") # plot forecasted values
Box.test(IPRforecast,lag = 5,type="Ljung-Box") # Apply the Ljung-Box
```